\documentclass[sigplan,nonacm]{acmart}
\usepackage{url}
\usepackage{xspace} 
\usepackage{subfig}
\usepackage{enumitem}
\usepackage{tikz}
\usepackage{multirow}
\usepackage{diagbox}
\usepackage{hyperref}
\usepackage{amsmath} 

\usepackage{caption} 
\newcommand{\ev}{\textit{CacheX}\xspace}
\newcommand{\ve}{\textsc{vev}\xspace}
\newcommand{\vs}{\textsc{vscan}\xspace}
\newcommand{\vc}{\textsc{vcol}\xspace}
\newcommand{\vt}{\textsc{vtop}\xspace}
\newcommand{\vcpc}{\textsc{cap}\xspace}
\newcommand{\lcas}{\textsc{cas}\xspace}

\newcommand{\lb}{\textsc{l2fbs}\xspace}
\newcommand{\rdtsc}{\textsc{rdtsc}\xspace}

\newcommand{\cpuid}{\textsc{cpuid}\xspace}

\usepackage{tikz} 
\newcommand*\whitecircled[1]{%
  \tikz[baseline=(char.base)]{
    \node[draw=black, fill=white, circle, inner sep=0.2pt, minimum size=1em, font=\bfseries\footnotesize] (char) {\textcolor{black}{#1}};
  }%
}

\begin{document}
\pagestyle{plain}
\title{Optimizing CPU Cache Utilization in Cloud VMs with Accurate Cache Abstraction}

\author{Mani Tofigh}
\affiliation{
  \institution{Hofstra University}
  \city{Hempstead}
  \state{NY}
  \country{USA}
}
\email{MTofigh1@pride.hofstra.edu}
\author{Edward Guo}
\affiliation{
  \institution{Columbia University}
  \city{New York}
  \state{NY}
  \country{USA}
}
\email{eg3415@columbia.edu}
\author{Weiwei Jia}
\affiliation{
  \institution{University of Rhode Island}
  \city{Kingston}
  \state{RI}
  \country{USA}
}
\email{weiwei.jia@uri.edu}
\author{Xiaoning Ding}
\affiliation{
  \institution{New Jersey Institute of Technology}
  \city{Newark}
  \state{NJ}
  \country{USA}
}
\email{dingxn@njit.edu}
\author{Zirui Neil Zhao}
\affiliation{
  \institution{University of Texas at Austin}
  \city{Austin}
  \state{TX}
  \country{USA}
}
\email{neil.zhao@utexas.edu}
\author{Jianchen Shan}
\affiliation{
  \institution{Hofstra University}
  \city{Hempstead}
  \state{NY}
  \country{USA}
}
\email{jianchen.shan@hofstra.edu}

\begin{abstract}

This paper shows that cache-based optimizations are often ineffective in cloud virtual machines (VMs) due to limited visibility into and control over provisioned caches. In public clouds, CPU caches can be partitioned or shared among VMs, but a VM is unaware of cache provisioning details. Moreover, a VM cannot influence cache usage via page placement policies, as memory-to-cache mappings are hidden. The paper proposes a novel solution, CacheX, which probes accurate and fine-grained cache abstraction within VMs using eviction sets without requiring hardware or hypervisor support, and showcases the utility of the probed information with two new techniques: LLC contention-aware task scheduling and virtual color-aware page cache management. Our evaluation of CacheX's implementation in x86 Linux kernel demonstrates that it can effectively improve cache utilization for various workloads in public cloud VMs.

\end{abstract}


\maketitle

\section{Introduction}

CPU caches play a crucial role in accelerating data access, and many optimizations have been developed to exploit their benefits. However, little attention has been devoted to evolving these optimizations for CPU caches available in virtual machines (VMs), which serve as the building blocks of cloud infrastructures. For simplicity, we refer to CPU caches accessible by VMs as vCache throughout this paper.

One key challenge lies in how vCache is provisioned to VMs. CPU caches, like the last-level cache (LLC), are often shared among co-located VMs. When such sharing leads to inter-VM cache interference, hardware features like Intel Cache Allocation Technology (CAT)~\cite{intelcat} and software approaches like page coloring~\cite{zhong2024managing, kessler1992page, zhang2009towards} can partition the cache among VMs. To avoid cache under-utilization, dynamic partitioning strategies~\cite{funaro2016ginseng, shahrad2021provisioning, xu2018dcat, xiang2018dcaps} are proposed to match cache provisions with the demands of VM workloads. However, the hypervisor does not expose vCache provisioning details, particularly its contended and dynamic nature, to VMs. Another major challenge lies in how memory is provisioned to VMs. Guest physical pages are backed by host physical pages, and their mappings are hidden from VMs. As a result, a VM is unaware of how its guest physical addresses (GPAs) map to cache sets, as that is determined by the underlying host physical addresses (HPAs). 

Due to these challenges, the existing vCache abstraction fails to accurately depict key cache characteristics. For instance, hypervisors often expose vCache as if it were identical to the host’s caches, assuming fixed size and associativity without any contention. Moreover, a page’s cache color, the specific set of cache sets it maps to, is often mistakenly inferred from its GPAs rather than its actual HPAs. This opacity makes cache-based optimizations ineffective.

One approach to address that is with hardware support. vLLC~\cite{kim2015vcache} extends TLB entries to store GPAs for LLC indexing, enabling a VM to manage LLC independently with effective color-based page placement. However, such architectural support won't be available in commodity hardware any time soon. Another approach is to rely on hypervisor support. For instance, CoPlace~\cite{shang2021coplace} modifies the hypervisor’s page allocation to ensure that guest physical pages with different GPA-derived colors map to host physical pages with distinct HPA-derived colors. This alignment makes color-based page placement more effective within the VM. CacheSlicer~\cite{shahrad2021provisioning} enables users to specify the desired number of LLC ways allocated to a cache-sensitive VM. vCAT~\cite{xu2017vcat} exposes the virtualized cache partitions to VMs from the hypervisor, allowing each VM to further divide the partitions among tasks. These work either pass hints from VMs to the host or expose cache information from the host to VMs—both of which necessitate hypervisor modifications. However, in the multi-cloud era~\cite{yang2023skypilot}, adopting these modifications is difficult due to the diverse or closed-source nature of hypervisors. Moreover, relying on hints from VMs can cause conflicts when their optimization goals differ, and exposing more
host-internal information to VMs raises security risks.

Thus, we aim to evolve cache-based optimizations \textbf{with} accurate vCache abstraction and page colors \textbf{within} the VM, which has the best knowledge of the workload, \textbf{without} requiring hardware support or hypervisor modifications. This approach can avoid the limitations of the previous two approaches and enables VMs to independently enhance cache efficiency for their specific workloads in multi-cloud environments, without relying on cloud providers to adopt new hardware or hypervisor-level remedies. To achieve this goal, we made the following contributions. 

Our \underline{first} contribution (§~\ref{sec: background}) is to systematically analyze and demonstrate through experiments on a commodity system (x86 Linux VMs on KVM~\cite{kivity2007kvm} hypervisor) the major impacts of inaccurate vCache abstraction and page colors. First, due to opaque cache provisioning, the vCache sizes reported to a VM often do not reflect the actual capacity it can utilize. This makes cache size-based optimizations ineffective and can lead to counterproductive effects. For instance, the scheduler may avoid migrating tasks to preserve cache affinity, even when migrating them to virtual CPUs (vCPUs) backed by LLCs with larger effective capacity can improve performance. Second, incorrectly relying on GPAs to infer page colors prevents page coloring techniques, such as pollute buffer~\cite{soares2008reducing}, from effectively reducing intra-VM cache conflicts. Third, vCache associativity is often assumed static and uniform, while in reality it varies with cache partitioning, and the effective per-set capacity changes dynamically under contention. Inter-VM cache interference could be reduced if VMs had access to per-set contention information, allowing page placement to favor less-contended cache sets.

Our \underline{second} contribution (§~\ref{sec: evcache}) is to design \ev, a novel solution to probe accurate and fine-grained vCache abstraction using eviction sets~\cite{liu2015last, vila2019theory} within the VM. The associativity of a vCache set is determined by the size of its minimal eviction set---the minimal set of memory blocks that can fully occupy the vCache set upon access. By accessing this eviction set, waiting for a specified time window, and re-accessing it, the eviction rate (i.e., the percentage of lines evicted per millisecond) of a vCache set is measured to reveal its effective capacity and contention from co-located VMs. Furthermore, since pages with different HPA-derived colors map to disjoint sets of cache sets, they can be grouped by testing whether they are evicted by the same minimal eviction set, called a color filter. Each group is assigned a \textit{virtual} color. To monitor dynamic contention with low overhead, only the eviction rates of a representative subset of vCache sets are periodically probed and aggregated to infer per-LLC and per-color contention information. To further reduce probing cost, eviction set construction, color filtering, and contention monitoring are all parallelized. To the best of our knowledge, \ev is the first to apply eviction sets for vCache probing within cloud VMs.

Our \underline{third} contribution (§~\ref{sec: casestudy}) is to showcase how \ev can be utilized to improve vCache utilization in VMs. We propose two new CacheX-based techniques: LLC contention-aware task scheduling (\lcas) and virtual color-aware page cache management (\vcpc), demonstrating how \ev can enhance existing cache-based optimizations. Particularly, \lcas encourages task migrations to prioritize idle vCPUs with less-contended LLC and avoid counterproductive cache affinity. \vcpc extends SRM-Buffer~\cite{ding2011srm}, a page coloring technique, to steer page cache accesses toward more-contended LLC sets, simultaneously reducing intra-VM cache pollution and absorbing inter-VM interference.

Our \underline{fourth} contribution (§~\ref{Sec: implementation}) is to implement \ev and the CacheX-based techniques in x86 Linux VM. \ev is realized via three microbenchmarks: \ve, \vc, and \vs. \ve adapts the state-of-the-art approach~\cite{zhao2024last} for cloud VMs by enhancing timing accuracy and incorporating vCPU topology awareness via integration with \vt~\cite{vsched}. It constructs minimal L2 and LLC eviction sets: the L2 ones serve as color filters used by \vc to classify free pages into colored free page lists, while the LLC ones are used by \vs to determine LLC associativity and monitor representative LLC sets for per-LLC and per-color contention levels, which are then leveraged by \lcas and \vcpc, respectively. \lcas extends \texttt{scx\_rusty}~\cite{rusty} by modifying its CPU selection heuristics to prioritize vCPUs with less-contended LLC. \vc performs free page allocation and color filtering in user space, while its kernel component organizes these pages into colored lists consumed by \vcpc in kernel's page cache allocation.

Our \underline{final} contribution (§ \ref{sec: evaluation}) is to evaluate \ev and the CacheX-based techniques with diverse workloads in public cloud VMs. We first show that eviction set construction, color filtering, and eviction rate monitoring are successfully accelerated through parallelization. To validate correctness, we perform sanity checks in local VMs using a custom hypercall that exposes GPA-to-HPA mappings, confirming that page colors are accurately identified and that constructed LLC eviction sets can monitor representative cache sets. We further show that eviction rate probing is accurate and that the default wait window reliably detects contention. Through experiments on public cloud VMs, we reveal that dynamic, asymmetric LLC contention and skewed page color distributions are common. Finally, we evaluate each CacheX-based technique: \lcas effectively steers tasks toward less-contended LLCs, and \vcpc reduces both intra- and inter-VM cache interference. We also show that \ev evolves cache-based optimizations in VMs at minimal cost, introducing only slight overhead from periodic probing.

\section{Background and Motivation}
\label{sec: background}

\subsection{vCache Abstraction and Page Color}

VMs depend on accurate vCache abstraction and page color information to drive optimizations across operating systems, language runtimes, and applications. The guest kernel probes \cpuid emulated by the hypervisor to infer cache properties and relies on GPAs to determine page color. Both are often inaccurate in cloud VMs, leading to performance issues.

\subsubsection*{Mismatched vCache Size.} 
The L2 and LLC caches can be shared among VMs or partitioned using mechanisms such as page coloring or Intel CAT. However, guest VMs remain unaware of such sharing or partitioning. They lack visibility into the actual effective vCache sizes, which are determined by cache partitioning policies or contention from neighbor VMs. This mismatch in size is most pronounced in the LLC rather than the L2 cache, as cloud providers often avoid scheduling vCPUs of different VMs on sibling hardware threads to prevent L2 cache sharing for security reasons~\cite{corescheduling}. Due to the mismatch, cache size-based optimizations become ineffective. For instance, self-adjusting applications~\cite{song2021cacheinspector} may mis-modulate output quality to meet latency goals, and language runtimes like \texttt{glibc} may misconfigure non-temporal store thresholds~\cite{memcpycachesize, glibcmemcpythreshold}.

\subsubsection*{Unreliable Page Color.}
Memory is logically divided into blocks (typically 64 bytes), which are cached in cache lines of the same size. Caches are organized into multiple equally sized sets, and each memory block maps to a single cache set, enabling efficient lookup. The number of cache lines within a set defines the cache's associativity. On modern processors, the LLC sets are grouped into slices~\cite{soori2024nucalloc}.

\begin{figure}[h!]
    \centering    \includegraphics[width=0.79\columnwidth]{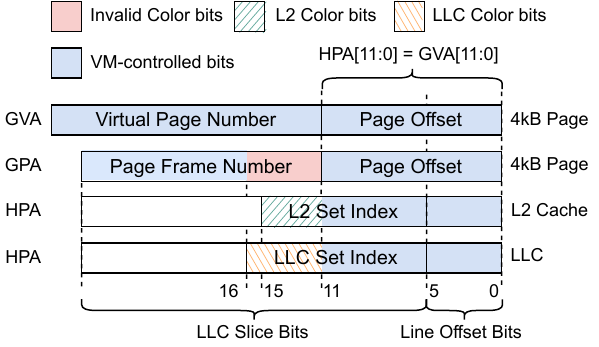}
    \caption{Mapping addresses to Skylake-SP’s L2 and LLC.}
    \label{fig:address-mapping}
\end{figure}

Figure~\ref{fig:address-mapping} illustrates the address mapping to L2 and LLC sets in a VM on our Intel Skylake-SP CPU, a model widely used in the cloud. Detailed cache parameters are provided in Table~\ref{tab:skylakecpu}. When a workload accesses a guest virtual address (GVA), the address is translated to a GPA, which is then mapped to an HPA by the hypervisor. The L2 cache uses HPA bits 15–6 as the set index bits to map an HPA to an L2 set, whereas the LLC mapping is a two-step process: (1) the HPA, excluding the lowest 6 bits, is hashed to determine the target LLC slice~\cite{mccalpin2021mapping}, and (2) HPA bits 16–6 index into a set within that slice. The low-order 12 bits of the HPA and GVA are identical for the standard 4 kB page size. We define two pages as having the same L2 cache color if their HPA bits 15–12 match, meaning their blocks map to the same set of L2 sets. Likewise, LLC colors are defined by matching HPA bits 16–12. Page coloring techniques use these color bits to control page placement, enabling either more even pressure on cache sets to reduce conflicts or explicit cache partitioning by allocating pages with distinct colors to different applications.

However, VMs are unaware of the GPA-to-HPA mappings and instead incorrectly rely on GPA bits 16–12 to infer cache color. This inference is unreliable, especially since GPA-to-HPA mappings are dynamic~\cite{shang2021coplace, bui2019extended} and can be changed by the hypervisor in response to events such as memory compacting and ballooning, none of which are visible to VMs. As a result, page coloring becomes ineffective: pages with the same GPA-derived color may map to disjoint sets.

\subsubsection*{Opaque Set Capacity} On a physical machine, cache associativity is an architectural property that determines the set capacity (i.e., the number of cache lines (ways) per set), and all ways are fully accessible to the system. In contrast, in cloud VMs, set capacity is often opaque due to two main reasons. First, cache sets may be way-partitioned among VMs using tools like Intel CAT, so their associativity is limited by the assigned cache ways. Second, cache partitions or the entire cache may be shared among VMs. Cloud providers might reserve dedicated partitions for cache-sensitive VMs~\cite{shahrad2021provisioning}, while assigning standard or low-cost VMs to shared partitions. Thus, the effective set capacity is influenced by contention from co-located VMs, and this contention may vary across sets. The asymmetry can stem from workload-specific cache access patterns or host memory fragmentation, where the hypervisor maps guest physical pages to non-contiguous host pages with skewed colors. This can place uneven pressure on certain sets, making the effective set capacity in a VM dynamic and asymmetric. If VMs were aware of per-set contention levels, they could avoid heavily contended sets to reduce inter-VM cache interference.

\subsection{Motivating Experiments}

We conduct experiments to demonstrate that fine-grained vCache abstraction and page color, if accurately exposed, can enhance existing cache-based optimizations in cloud VMs. Detailed benchmark descriptions and experimental settings can be found in the evaluation section (§~\ref{sec: evaluation}). 

\subsubsection*{Counterproductive Cache Affinity} 

The Linux scheduler employs cache affinity to place tasks on cores with warm caches, favoring the same core or LLC domain~\cite{lozi2016linux} (a group of cores that share the LLC) to reuse cached data and share data between threads efficiently. To preserve this cache affinity, the scheduler discourages task migrations across LLC domains~\cite{schedcache}. However, in a VM, this optimization can become counterproductive by preventing tasks from being moved away from a heavily contended LLC domain, even when idle vCPUs exist in other less-contended LLC domains. 

To demonstrate the impact of such unawareness of LLC asymmetry, we run a set of cache-sensitive workloads in a 16-vCPU VM, with its vCPUs evenly split across two LLC domains and pinned to 8 cores per domain. To create LLC asymmetry, we run \textit{cache polluter} on the remaining 12 cores in one LLC domain, accessing large memory regions with a 64 B stride to stress the LLC. In the other domain, we run the CPU-intensive sysbench~\cite{sysbench} on the remaining 12 cores to minimize LLC contention. Each cache-sensitive workload runs with 8 threads under two settings: one using the default Linux EEVDF~\cite{eevdf} scheduler, and the other with all threads pinned to the non-contended LLC domain.

\begin{figure}[t]
    \centering
    \subfloat[\centering EEVDF vs. Pinned]{{\includegraphics[width=0.45\columnwidth]{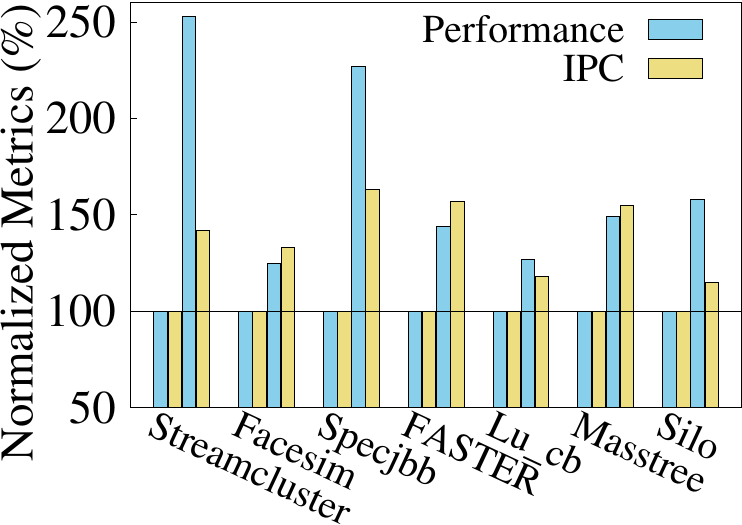} }}%
    \qquad
    \subfloat[\centering Specjbb under EEVDF]{{\includegraphics[width=0.446\columnwidth]{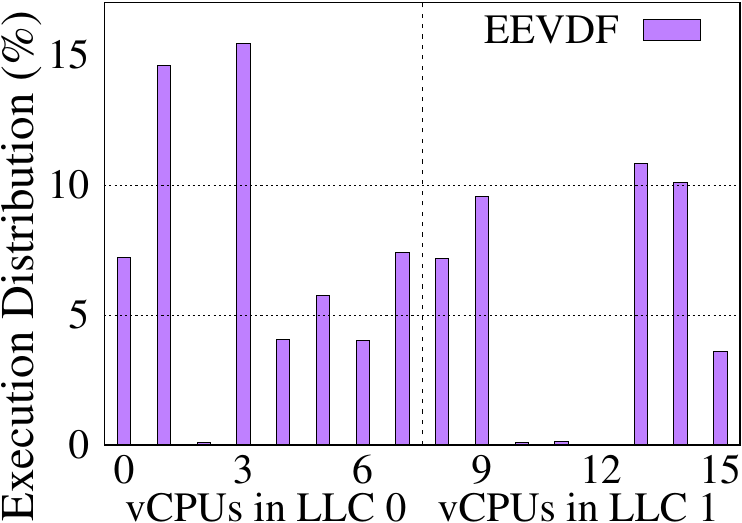} }}%
    \caption{\small The impact of counterproductive cache affinity. In (a), the metrics are normalized to EEVDF, and the first two bars represent the metrics under EEVDF, while the last two bars depict those when threads are pinned. In (b), LLC 0 is heavily contended.}%
    \label{fig:badcacheaffinity}%
\end{figure}%

As shown in Figure~\ref{fig:badcacheaffinity}a, \textit{pinning} significantly increases instructions per cycle (IPC) (averaging 38\%) and improves workload performance (averaging 73\%), compared to EEVDF. We utilize IPC to indirectly demonstrate cache efficiency, as cache miss profiling tends to be less reliable in VMs~\cite{song2021cacheinspector}. Figure~\ref{fig:badcacheaffinity}b, using specjbb~\cite{specjbb} as an example, shows that under EEVDF, specjbb is constrained to the polluted LLC domain for 59\% of its runtime. This observation highlights the importance of considering the dynamic LLC contention in vCPU selection for task placement (§~\ref{sec: lcas}).

\subsubsection*{Ineffective Page Coloring}
Page coloring is a widely used software technique for cache partitioning~\cite{zhang2009towards}, applied to isolate cache polluters~\cite{soares2008reducing, ding2011srm, park2020page}, mitigate contention in tiered memory systems~\cite{zhong2024managing}, and defend against cache-based side channels~\cite{shi2011limiting, hofmann2024gaussian, volos2024principled}. To illustrate how unreliable GPA-derived page color undermines this technique, we create a 16-vCPU VM that co-runs a cache-sensitive workload alongside \textit{cache polluter}, each using 8 vCPUs pinned to separate cores within the same LLC domain. To minimize interference, we adopt an optimization inspired by the pollute buffer design~\cite{soares2008reducing}, restricting \textit{cache polluter} to accessing pages of a single LLC color. We implement two variants: one selects pages based on GPA color bits, and the other uses HPA color bits (retrieved via a custom hypercall). We evaluate both variants under two memory conditions: \textit{contiguous}, where guest memory is mostly backed by contiguous host pages when the VM is freshly booted; and \textit{fragmented}, where guest memory maps to more non-contiguous host pages after repeated kernel compilations in the aged VM.

\begin{figure}[t]
    \centering
    \subfloat[\centering Pollute buffer isolation]{{\includegraphics[width=0.46\columnwidth]{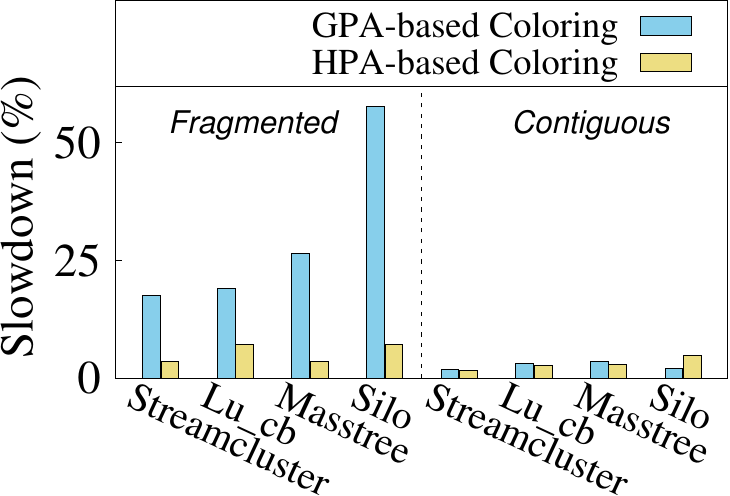} }}%
    \qquad
    \subfloat[\centering Page color used by polluter]{{\includegraphics[width=0.436\columnwidth]{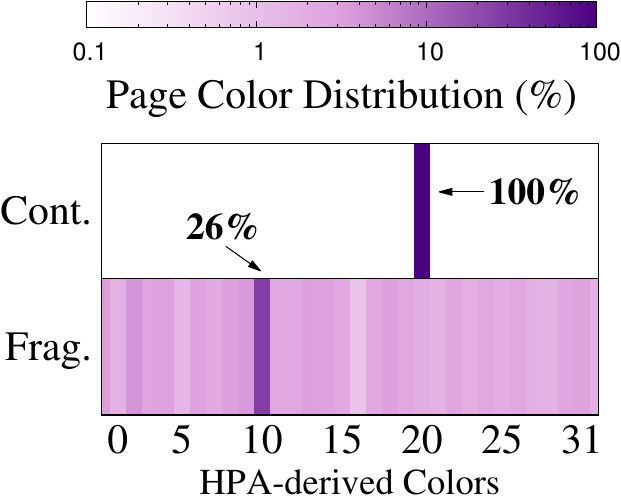} }}%
    \caption{\small (a) Slowdowns caused by \textit{cache polluter} relative to solo runs. (b) GPA-derived color's distribution in HPA-derived colors.}%
    \label{fig:invalidcolor}%
\end{figure}

Figure~\ref{fig:invalidcolor}a shows that \textit{cache polluter}, when isolated using GPA-based page coloring, still slows down the workload by 29.1\% on average under fragmented mode. As revealed in Figure~\ref{fig:invalidcolor}b, memory fragmentation causes pages with the same GPA-derived color used by \textit{cache polluter} to be mapped to a broader range of HPA-derived colors, leading to increased unintended cache interference. In contrast, contiguous mode and HPA-based coloring provide effective isolation. Since guest memory is often non-contiguous in public clouds~\cite{fakehuge}, having accurate page color information is essential for effective page coloring optimizations (§~\ref{sec: vcpc}).

\subsubsection*{Avoidable Set Contention} 
\begin{figure}[h!]
    \centering
    \includegraphics[width=1\columnwidth]{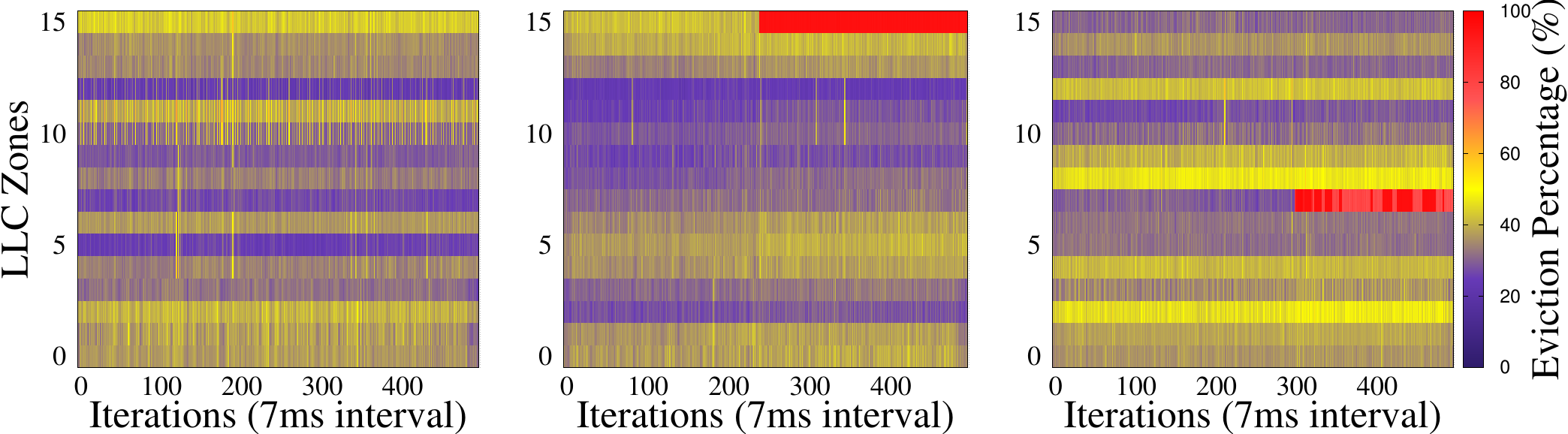}
    \caption{\small Avoidable inter-VM cache conflicts under asymmetric contention across LLC sets.}
    \label{fig:hotzoneconflict}
\end{figure}

Some sets can become \textit{hotter} as they experience more contention than others due to uneven pressure from co-located VMs. To demonstrate the impact, we create two 10-vCPU VMs with their vCPUs pinned to 20 cores in the same LLC domain. One VM runs nginx~\cite{reese2008nginx}, while the other runs \textit{cache poisoner} that continuously accesses memory pages mapped to just 1/16 of the LLC sets. Figure~\ref{fig:hotzoneconflict} shows the activity of 16 LLC zones over time. Each vertical line represents the eviction activity of 128 monitored sets in a zone, measured by \vs (§~\ref{sec: vset}). A higher eviction percentage, indicated in red, suggests that more lines were evicted during the interval. When nginx runs alone (Figure~\ref{fig:hotzoneconflict}-left), some zones are hotter due to its own access pattern. However, when the poisoner intentionally stresses one of these hot zones (Figure~\ref{fig:hotzoneconflict}-middle), nginx throughput drops by 9.4\%. In contrast, when the poisoner hits the cold zone (Figure~\ref{fig:hotzoneconflict}-right), the performance degradation is eliminated. 

This result highlights the importance of incorporating per-set contention awareness into guest page placement policies to reduce inter-VM cache interference (§~\ref{sec: vcpc}).
\section{The CacheX Approach}
\label{sec: evcache}

We aim to profile fine-grained vCache abstraction and page color within the VM, but face two major challenges. First, performance monitoring units (PMUs), commonly used in cache profiling tools~\cite{abel2020nanobench}, are rarely available in public clouds~\cite{song2021cacheinspector}. Even when virtual PMUs are enabled in cloud VMs, they are often unreliable~\cite{vPMU} and expose only a limited set of events~\cite{vpmuuncore}. Second, virtualization adds significant profiling interference due to factors like misaligned huge pages~\cite{fakehuge}, dynamic vCPU topology~\cite{vsched}, and unknown cache architectures (e.g., inclusive vs. non-inclusive caches~\cite{yan2019attack}).

To overcome these challenges, we introduce \ev, a vCache probing technique that operates entirely within the VM, requiring no hardware or hypervisor support. The key idea is to use eviction sets for accurate and lightweight probing. An eviction set consists of memory blocks that fully occupy a single cache set; a \textit{minimal} eviction set is one where removing any block breaks this property. The addresses of the blocks in such a set are considered \textit{congruent}~\cite{vila2019theory}. By detecting the size of the minimal eviction set, \ev can reveal vCache associativity. By priming (accessing) and probing (re-accessing) the eviction set later, \ev counts evictions caused by co-located VMs, providing a measure of per-set contention. Last, \ev leverages the observation that pages with the same color are evicted only by eviction sets mapped to cache sets of the same color. By filtering pages using eviction sets of all colors, \ev can group them by color within the VM.

\begin{figure}[h!]
    \centering    \includegraphics[width=0.87\columnwidth]{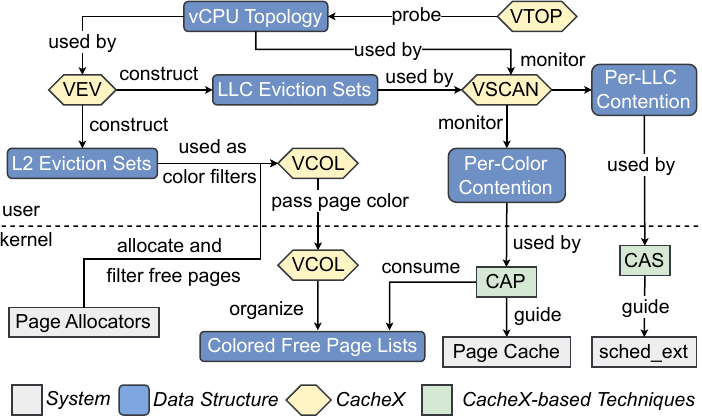}
    \caption{CacheX Approach Overview.}
    \label{fig:evcacheoverview}
\end{figure}

Figure~\ref{fig:evcacheoverview} presents an overview of \ev, including its user- and kernel-level components and their interactions with other elements within the VM. \ve constructs minimal L2 and LLC eviction sets, which are used by \vc to filter pages and build colored free page lists, and by \vs to infer LLC associativity and monitor LLC sets for per-LLC and per-color contention. \vt~\cite{vsched} is integrated to assist in LLC eviction set construction and per-LLC monitoring. We demonstrate the utility of \ev via two new techniques (§~\ref{sec: casestudy}): LLC contention-aware task scheduling (\lcas) and virtual color-aware page cache management (\vcpc).

\subsection{Eviction Set Construction}

\subsubsection*{Basic Steps} A naïve method to construct all minimal eviction sets involves repeating the following two steps at each \textit{aligned page offset} (i.e., those with GVA bits [5:0] = 0 and bits [11:6] varying, based on Figure~\ref{fig:address-mapping}): \whitecircled{1} Initialize a pool of candidate addresses at aligned page offset \textit{P}. The pool must be large enough to construct all minimal eviction sets at that offset. Its size, denoted as $P_s$, is calculated as follows: for the L2 cache, $P_s\!=\!W\!\times\!2^{N_{UI}}\!\times\!C$; for the LLC,  $P_s\!=\!W\!\times\!2^{N_{UI}}\!\times\!N_{slices}\!\times\!C$. Here, \textit{W} is the number of cache ways, $N_{UI}$ is the number of uncontrollable set index bits, $N_{slices}$ is the number of LLC slices, and \textit{C} is a scaling factor (set to 3 in our construction) to account for uneven address distribution across cache sets and slices. \whitecircled{2} Remove a \textit{target} address from the pool. If it cannot be evicted by previously-built minimal eviction sets at offset \textit{P}, perform \textit{pruning}: iteratively exclude a candidate from consideration if the remaining ones in the pool can still evict the target. This continues until a new minimal eviction set is found, whose addresses are then removed from the pool. The process repeats until the maximum number of minimal eviction sets at offset \textit{P} is built or the pool is empty.

\subsubsection*{State of the Art} Eviction sets are primarily used in side-channel attacks~\cite{irazoqui2015s, liu2015last, yan2019attack}. Due to shorter attack windows and larger caches, recent work focuses on speeding up eviction set construction by filtering candidate pools~\cite{zhao2024last, morgan2025slice+}, accelerating pruning~\cite{vila2019theory, purnal2021prime+, zhao2024last}, and bulk construction~\cite{kessous2024prune+}. Fast construction is crucial to reducing probing overhead in our case. Among existing methods, \ve adopts the approach proposed by Zhao et al.~\cite{zhao2024last}, because it is resilient to noisy cloud environments and has been evaluated on real cloud CPUs. We refer to it as \lb. \lb leverages memory-level parallelism (MLP) to quickly test candidate eviction sets, reducing false positives that occur when the target is evicted due to cache activity from other tenants, and uses binary search to efficiently prune candidates. To further avoid noise during LLC eviction set construction, before the pruning stage, \lb first filters out from the pool any address that cannot be evicted by the target’s L2 eviction set, since only those matching the target’s L2 set index bits (part of the LLC set index bits) and thus evictable in the L2 cache can potentially be congruent with the target in the LLC. In addition, \lb supports both inclusive and non-inclusive caches, and doesn't rely on specific replacement policies or huge pages.

\subsubsection*{Adapting to Cloud VMs} Since \lb was originally developed for the Function-as-a-Service (FaaS) platform built on containers~\cite{googlfaas}, we made two key adaptations for cloud VMs. First, latency measurements used to determine where a target is cached become unreliable in VMs. Although \lb preloads the TLB entry by triggering address translation without memory access to avoid costly two-dimensional page walks~\cite{merrifield2016performance} during measurements, we still observe latency spikes even when the target resides in L1/L2 caches. Our experiments revealed that this was caused by unstable guest TSC (Time-Stamp Counter)~\cite{guesttsc} readings via \rdtsc~\cite{tscscaling}. We mitigate this by warming up the timer with dummy \rdtsc calls before measurements. Second, LLC eviction set construction suffers from high overhead or fails due to the opaque and dynamic vCPU topology in VMs. As in prior work~\cite{yan2019attack, purnal2021prime+}, \lb uses a helper thread paired with a construction thread to access a target simultaneously, pulling it to the LLC by changing the cache line state to \textit{shared}. This improves success rates compared to the single-thread version, which relies on eviction from lower-level caches to move the target into the LLC. However, this method only works when the two threads run on vCPUs that are in the same LLC domain but are not SMT siblings (topological information that is hidden from the guest). To enable this, we integrate \vt~\cite{vsched}, which periodically infers accurate vCPU topology by measuring inter-vCPU latencies without hardware or hypervisor support. It is worth noting that \vt cannot infer the vCPU-to-core mapping, which remains hidden from VMs unless exposed through paravirtualized interfaces~\cite{liu2023cps}. However, this mapping is essential to \textit{slice filtering}~\cite{morgan2025slice+}, which filters out addresses that do not map to the same LLC slice as the target. Therefore, we are unable to adopt this technique.

\subsection{Page Color Identification}

\subsubsection*{Constructing Color Filters}
Although the exact color bits are hidden, we can group pages based on whether they are evicted by eviction sets mapped to cache sets of the same color. Each group can be assigned a virtual color, sufficient for enabling page coloring. To achieve this, \ve builds minimal L2 eviction sets, called \textit{color filters}, to identify pages of all L2 colors. It begins with a pool of candidate addresses at page offset \texttt{0x0}. This pool is large enough to build minimal eviction sets across all L2 sets at that offset. Unless the host enforces page coloring to prevent a VM from receiving pages in all colors, on our Skylake-SP CPU, \ve can build up to 16 color filters used by \vc to organize free pages into lists. This enables page coloring techniques to partition the L2 cache and LLC into 16 zones. It is worth noting that building filters for different LLC colors is infeasible. Since the slice bits are uncontrollable, we cannot ensure that all minimal LLC eviction sets built at the same page offset reside in the same slice. As a result, while two minimal L2 eviction sets at the same page offset are guaranteed to have distinct L2 colors, two minimal LLC eviction sets at the same page offset may share the same LLC color but map to different slices, making reliable LLC color filtering impractical.

\subsubsection*{Parallel Filtering}
Unlike color identification via HPA, which is instantaneous, identifying color via GVA is time-consuming. In the worst case, a page must be tested against all color filters before its virtual color can be determined. To speed this up, we propose \textit{parallel color filtering}, which tests a page against all color filters simultaneously. Although each color filter is initially constructed at page offset \texttt{0x0}, it can be replicated to any of the other 63 aligned page offsets. On our Skylake-SP CPU, we shift the 16 color filters to different page offsets. This enables us to select 16 addresses from a page, each at a different offset, and test them in parallel against the filters at matching offsets, without interference. Only one address is evicted, revealing the page’s color. 

\subsection{Set Associativity and Contention Probing}
\label{sec: vset}

\vs probes LLC associativity and set-level contention caused by co-located VMs. The associativity is determined by the size of minimal LLC eviction sets built by \ve. Contention is monitored by periodically priming and probing these eviction sets while pausing the VM's own workloads, revealing cache activity introduced by other tenants. While the probed associativity helps infer the LLC capacity available to a VM in the absence of contention, estimating effective capacity requires accounting for both the VM’s own cache usage and external contention when present. Nonetheless, set-level contention alone is sufficient to guide critical decisions such as task scheduling (§~\ref{sec: lcas}) and page placement (§~\ref{sec: vcpc}).

\begin{figure}[h!]
    \centering    \includegraphics[width=0.64\columnwidth]{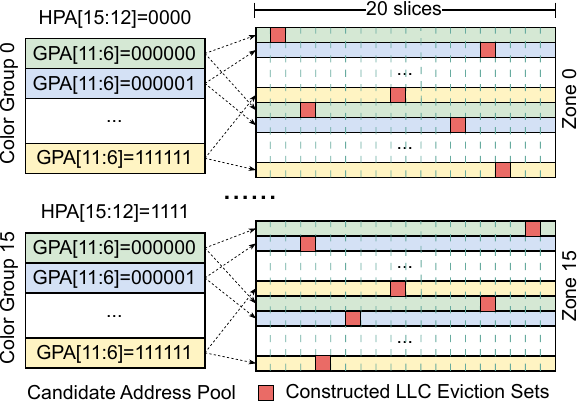}
    \caption{Parallel Eviction Set Construction.}
    \label{fig:vsetparabuild}
\end{figure}

\subsubsection*{LLC Sets to Probe} While probing all LLC sets offers complete visibility, constructing and monitoring minimal eviction sets for all of them is prohibitively expensive (§ ~\ref{sec: vevictev}). To address this challenge, \vs probes only one set per LLC set index. Since addresses with the same set index are evenly distributed across slices~\cite{mccalpin2021mapping}, contention observed in a single set is representative of all sets sharing that index. To implement this efficiently, we introduce \textit{parallel eviction set construction} (Figure~\ref{fig:vsetparabuild}), illustrated using our Skylake-SP CPU. The LLC is visualized as a 2D grid: each \textit{row} corresponds to sets with the same index across all slices, and each \textit{column} to sets within one slice across different indices. \ve first initializes a candidate pool large enough to build all minimal LLC eviction sets. It then applies color filters to split it into 16 color groups and partitions each group by aligned page offset. From each offset partition, it constructs $f\!=\!2$ minimal eviction sets to target two sets in distinct rows. In practice, we set $f\!=\!4$ (§~\ref{sec: vsetev}) to lower the risk that the constructed eviction sets share the same set index due to the uncontrollable HPA bit 16. Constructions on offset partitions are assigned to multiple pairs of threads (constructor and helper), each operating in parallel on disjoint \textit{rows} to avoid interference.

\subsubsection*{Set Contention Monitoring} 

To quantify contention in an LLC set, \vs uses \textit{windowed Prime+Probe}~\cite{purnal2021prime+} to measure how many cache lines are evicted by co-located VMs within a specified time window. Higher contention implies that less of the VM’s own data can persist in the set, leading to more cache misses. Compared to \textit{windowless Prime+Probe}~\cite{purnal2021prime+}, which mainly tracks access frequency, the windowed variant better reflects cache occupancy. For example, a VM that frequently touches a single line occupies only one way in a cache set, making frequency alone unreliable for gauging contention. Unlike side-channel attacks that require low-latency prime and probe phases to avoid missing victim accesses~\cite{purnal2021prime+, zhao2024last}, our contention probing exploits MLP~\cite{zhao2024last} to accelerate priming while performing the probe phase sequentially to measure access latency of each primed address to detect evictions accurately. In addition, we probe in reverse order to reduce self-evictions~\cite{liu2015last}. 

Choosing an appropriate window is key: if too long, all cache lines across all sets may be evicted; if too short, evictions may not occur. A default 7 ms window is empirically determined (§~\ref{sec: vsetev}) to strike a balance, capturing comparable evictions across sets. \vs also auto-adjusts the window: it shrinks when full eviction is observed across the sets and resets to default when evictions are absent. However, VM’s workloads are paused during prime and probe phases, which include this window, introducing non-trivial overhead. To mitigate this, \vs parallelizes monitoring by launching multiple thread pairs (monitor and helper), each handling a portion of the target sets. Each thread pair primes all assigned minimal eviction sets, waits, and then probes them. These optimizations reduce total monitoring time to under 10 ms (§~\ref{sec: vsetev}). Monitoring is conducted periodically (e.g., every second) to further reduce the overhead while ensuring prompt updates of set contention. Contention is normalized using the eviction rate (the percentage of lines evicted per millisecond) and smoothed using an exponentially weighted moving average (EWMA). Finally, \vs aggregates contention by LLC and by color for \lcas (§~\ref{sec: lcas}) and \vcpc (§ ~\ref{sec: vcpc}).
\section{Case Study}
\label{sec: casestudy}

This section demonstrates how CacheX enhances existing cache-based optimizations within cloud VMs.

\subsection{LLC Contention-Aware Task Scheduling}
\label{sec: lcas}

To demonstrate that \ev can enhance cache-aware scheduling, we introduce \lcas, an LLC contention-aware task scheduling optimization designed to address the issue of \textit{counterproductive cache affinity} in cloud VMs with asymmetric LLCs. We implement \lcas in \texttt{scx\_rusty}~\cite{rusty}, rather than EEVDF, to isolate its effects from complex scheduling heuristics. \texttt{scx\_rusty} is a hybrid BPF/user-space scheduler~\cite{schedext} that groups cores by LLC domain. Within each domain, tasks are dispatched from a dispatch queue to cores; cross-domain load balancing is guided from user space. Its task placement prioritizes a task’s previous core or the same domain as dependent threads (e.g., co-locating waker and wakee). To prevent tasks from being confined to a heavily-contended LLC domain when the system is underloaded and there are less-contended domains, \vs is leveraged to incorporate LLC contention awareness. Using per-LLC eviction rates reported by \vs, \lcas classifies domains into qualitative tiers, where domains with lower eviction rates are ranked higher. During task placement, idle vCPUs in higher-ranked domains are preferred. Since tasks are periodically preempted and re-dispatched by scheduler ticks, \lcas gradually steers them into less-contended domains. In addition, \lcas restricts load balancing from pulling tasks from a less-contended to a more-contended domain unless the source domain is saturated with high CPU utilization. To prevent tasks from being frequently switched among domains due to transient contention changes, a domain’s tier is only updated if its eviction rate consistently increases or decreases across three consecutive \vs monitoring intervals.

\subsection{Virtual Color-Aware Page Cache Management}
\label{sec: vcpc}

To showcase \ev's support for in-VM page coloring, we present \vcpc, a virtual color-aware page cache optimization. \vcpc follows the principle of SRM-Buffer~\cite{ding2011srm} to reduce LLC pollution from page cache accesses, which often exhibit poor temporal locality. For example, file-scanning workloads access data buffered in page cache sequentially, benefiting little from the LLC but evicting workload data with frequent reuse, thereby degrading performance. To mitigate this, upon page cache misses, \vcpc allocates pages from \vc's colored free page lists. Rather than limiting allocation to a single color, which constrains allocatable memory space and may increase page cache misses, \vcpc proceeds to the next color after exhausting pages in the current one. This ensures page cache accesses primarily impact a single LLC zone at a time, similar to SRM-buffer, reducing pollution without the drawbacks of fixed color allocation. Allocated pages are marked as non-movable~\cite{gfpflags} to preserve their colors.

Unlike SRM-buffer, \vcpc ranks the colored free page lists based on per-color contention measured by \vs, prioritizing pages with highly contended colors. This strategy directs page cache traffic toward \textit{hot} LLC zones (zones corresponding to colors with heavy contention), thereby effectively absorbing interference from co-located VMs that would otherwise evict workload data with strong locality but are less harmful to page cache data with low temporal reuse. A key challenge lies in adapting to dynamic contention: actively used page cache pages may end up in less-contended zones, while inactive ones remain in hot zones, undermining the intended optimization. To address this, \vcpc classifies colors into qualitative tiers based on their eviction rates. If the previously hottest color is in a lower rank than the new hottest color for three consecutive \vs monitoring intervals, \vcpc reclaims all file-backed page cache pages. This allows subsequent allocations, most likely used to buffer the actively accessed files, to use now-hotter colors. Effectively, page cache pages are recolored to absorb the shifting interference without significantly increasing page cache misses. 
\section{Implementation Details}
\label{Sec: implementation}

We rewrite \vt in C for direct embedding into \ve, with its topology propagation optimized by skipping checks that cannot aid vCPU distance inference. \ve is integrated into both \vs and \vc to build the required eviction sets. Since \vs and \vc operate mainly in user space, they are assigned to a high-priority cgroup to avoid interference from VM workloads. \lcas extends \texttt{scx\_rusty}, and \vs reports per-LLC contention to \lcas via a BPF map~\cite{bpfmap}, which guides task domain preference. \vcpc is implemented in Linux kernel v6.15.2. \vc allocates pages for color filtering in user space and assigns virtual colors via a custom kernel module before freeing them. \vc's kernel part then intercepts page free paths to insert colored pages, including those freed from page cache, into colored free page lists, and \vcpc modifies page cache allocation paths (e.g., \texttt{filemap\_alloc\_folio\_noprof}) to allocate from these lists. In addition to parallel color filtering, \vc parallelizes both user-space page allocation and kernel-level list insertion. Overall, \ev has 8,858 lines of code (LoC), with 271 LoC in the kernel. In user space, it incorporates 1,426 LoC from \lb functions and 886 LoC for \vt. \lcas contributes 125 LoC to \texttt{scx\_rusty} and \vcpc adds 238 LoC to the kernel.
\section{Evaluation}
\label{sec: evaluation}

Our evaluation aims to demonstrate the following:
\begin{itemize}[leftmargin=*]
    \item The performance, accuracy, and parameter sensitivity of \ev-\ve (§~\ref{sec: vevictev}), \vc (§~\ref{sec: vcolorev}), and \vs (§~\ref{sec: vsetev}).
    \item The dynamic, asymmetric LLC contention and skewed page color distribution probed in cloud VMs (§~\ref{sec:cloudprobe}).
    \item The effectiveness of \lcas (§~\ref{sec: lcasev}) and \vcpc (§~\ref{sec: vcpcev}) in cache interference reduction.
    \item The low overhead (§~\ref{sec: overhead}) incurred by \ev.
\end{itemize}

{
\renewcommand{\arraystretch}{0.96}
\begin{table}[h!]
\setlength{\tabcolsep}{2pt}
\footnotesize
\caption{\small Cache Parameters of Intel Gold 6138 CPU.}
\label{tab:skylakecpu}
\begin{tabular}{|l|l|}
\hline
\textbf{Structure} & \textbf{Parameters}                                  \\ \hline
L1                 & Data/Instruction: 32 kB, 8 ways, 64 sets, 64 B line  \\ \hline
L2                 & 1 MB, 16 ways, 1024 sets, non-inclusive to L1        \\ \hline
LLC Slice          & 1.375 MB, 11 ways, 2048 sets, non-inclusive to L1/L2 \\ \hline
Num Slices         & 20 slices                                            \\ \hline
\end{tabular}
\end{table}
}

\subsubsection*{Platforms} Our experiments use both local and cloud VMs, all running kernel v6.15.2. Local VMs allow us to measure the accuracy of \ev with host-side verification, and to control co-located VMs when evaluating \lcas and \vcpc. They are created with KVM on an HPE ProLiant DL580 Gen10 server (4 Intel Skylake-SP Gold 6138 20-core CPUs, 256 GB memory), with hyperthreading disabled and running the same kernel. Cache parameters are detailed in Table~\ref{tab:skylakecpu}. To observe LLC contention and page color distribution in the cloud, we probe three VMs from Google Cloud (N1 series)~\cite{googlevm}, AWS (M5dn series)~\cite{awsvm}, and Azure (Dsv2 series)~\cite{azurevm} with \ev. For comparability, all VMs are located in the \texttt{us-east} region and configured with 8 vCPUs (vCPU-to-core ratio set to 1) on CPUs with identical cache parameters (8-way L1, 16-way L2, and 11-way LLC, matching the local server). The Google VM is reconfigured with 10 vCPUs to evaluate \ve and with 40 vCPUs to examine asymmetric LLC contention.

\subsubsection*{Workloads}
To demonstrate that \ev can improve both throughput and latency for workloads across diverse domains, we select a suite of cache-sensitive benchmarks including PARSEC~\cite{bienia2008parsec} (canneal, ferret, facesim, lu\_cb, ocean\_cp), Tailbench~\cite{kasture2016tailbench} (specjbb, masstree, silo, moses, shore), Kernbench~\cite{kernbench}, DLRM~\cite{dlrmcode}, Pbzip2~\cite{pbzip2}, Nginx~\cite{reese2008nginx}, FASTER~\cite{fasterChandramouli2018}, and microbenchmarks (matmul, mergesort). Each experiment is executed ten times after warm-up runs, and averages are reported. We implement a \textit{cache polluter} to create LLC contention by accessing large memory regions with multithreading from co-located VMs, and a \textit{cache poisoner} built atop \vs to precisely stress specific LLC zones. Fio~\cite{fio} is used to generate cache pollution via page cache accesses.

\subsection{Performance and Accuracy of \ve}
\label{sec: vevictev}

{
\renewcommand{\arraystretch}{0.99}
\begin{table}[h!]
\setlength{\tabcolsep}{0.70pt}
\centering
\footnotesize
\caption{\small Minimal LLC Eviction Sets Construction.}
\label{tab: vevictev}
\begin{tabular}{|c|c|c|c|c|}
\hline
\textbf{VMs} & \textbf{Metrics} & \textbf{\lb (Full)} & \textbf{\ve (Full)} & \textbf{\ve ($\boldsymbol{f\!=\!4}$)} \\ \hline
\multirow{2}{*}{\begin{tabular}[c]{@{}c@{}}Local VM\\ (\scriptsize 10 vCPUs, 1 LLC)\end{tabular}} & Succ. Rate & 99.85\% & 99.96\% & 99.97\% \\ \cline{2-5} 
 & Time (s) & 45.05 ± 3.06 & 12.59 ± 3.42 & 2.24 ± 0.69 \\ \hline
\multirow{2}{*}{\begin{tabular}[c]{@{}c@{}}Local VM\\ (\scriptsize 20 vCPUs, 1 LLC)\end{tabular}} & Succ. Rate & 99.76\% & 99.92\% & 99.84\% \\ \cline{2-5} 
 & Time (s) & 44.79 ± 5.85 & 7.63 ± 2.13 & 1.05 ± 0.10 \\ \hline
\multirow{2}{*}{\begin{tabular}[c]{@{}c@{}}Local VM\\ (\scriptsize 20 vCPUs, 2 LLCs)\end{tabular}} & Succ. Rate & 46.57\% & 99.96\% & 99.91\% \\ \cline{2-5} 
 & Time (s) & 365.50 ± 312.29 & 8.64 ± 1.25 & 1.17 ± 0.09 \\ \hline
\multirow{2}{*}{\begin{tabular}[c]{@{}c@{}}Google VM\\ (\scriptsize 10 vCPUs, 1 LLC)\end{tabular}} & Succ. Rate & 99.78\% & 99.85\% & 99.85\% \\ \cline{2-5} 
 & Time (s) & 107.51 ± 2.27 & 31.08 ± 7.06 & 2.90 ± 0.24 \\ \hline
\end{tabular}
\end{table}
}

We compare \ve with \lb in both quiescent local VMs and a noisy cloud VM for constructing minimal L2 and LLC eviction sets. Each experiment is repeated 10 times locally and $1{,}000$ times in the cloud VM. \lb is patched with our guest TSC fix. Table~\ref{tab: vevictev} reports results for the LLC. In local VMs with 10 or 20 vCPUs pinned to a single LLC domain, \lb achieves high success rates and stable runtimes when building for all $40{,}960$ LLC sets. However, with 20 vCPUs spread across two LLC domains, the lack of vCPU topology awareness in \lb can cause its construction and helper threads to run on different domains, leading to long stalls, unstable runtimes, and frequent failures (46.57\% success rate).

{
\renewcommand{\arraystretch}{0.4}
\begin{table}[h!]
    \setlength{\tabcolsep}{2pt}
    \footnotesize
    \centering
    \caption{\small LLC Associativity Probed by \ve ($f\!=\!4$).}
    \begin{tabular}[t]{lccc}
    \toprule \textbf{Metrics} &\textbf{3 ways} &\textbf{5 ways} &\textbf{8 ways} \\
    \midrule
    Succ. Rate & 99.78\% & 99.82\% & 99.91\% \\
    \midrule
    Time (s) & 1.78 ± 0.05 & 1.74 ± 0.05 & 1.96 ± 0.09 \\
    \midrule
    Num Ways & 3.10 ± 0.32 & 5.40 ± 0.52 & 8.20 ± 0.42 \\
    \bottomrule
    \end{tabular}
    \label{tab: vevictcatev}
\end{table}%
}

\ve uses parallelization and integrates \vt, achieving $3.6\times$, $5.9\times$, and $42.3\times$ speedups over \lb in the three local VM configurations, while maintaining consistently high success rates. In the cloud VM (10 vCPUs in one LLC domain) with $57{,}344$ LLC sets, \ve is $3.5\times$ faster than \lb, with both maintaining high success rates. For the 4,096 minimal LLC eviction sets used by \vs ($f\!=\!4$), \ve builds them under 3s with 10 vCPUs and about 1s with 20 vCPUs, a negligible one-time cost. For the 16 color filters (minimal L2 eviction sets) used by \vc, both algorithms build them sequentially under 26 ms with a 100\% success rate. No duplication is observed in \vc's color filters. \vs's LLC eviction sets show negligible duplication (<1\%). We repeat the experiments on the 10-vCPU local VM while varying the LLC ways allocated via CAT. As shown in Table~\ref{tab: vevictcatev}, \vs accurately reports the LLC associativity, providing insight into the underlying CAT partition.

\subsection{Performance and Accuracy of \vc}
\label{sec: vcolorev}

{
\renewcommand{\arraystretch}{1.1}
\begin{table}[h!]
\setlength{\tabcolsep}{1pt}
\centering
\footnotesize
\caption{\small Colored Free Page Lists Construction Time (unit: ms).}
\label{tab: vcolorev}
\begin{tabular}{|c|c|c|c|c|}
\hline
\textbf{Settings} & \textbf{Methods} & \textbf{Allocation} & \textbf{Filtering} & \textbf{Insertion} \\ \hline
\multirow{2}{*}{\begin{tabular}[c]{@{}c@{}}20 vCPUs\\ 500 MiB\end{tabular}} & Seq. & 927.20 ± 11.69 & 2097.60 ± 16.55 & 4049.80 ± 233.54 \\ \cline{2-5} 
 & Para. & 190.00 ± 4.30 & 164.00 ± 2.45 & 665.00 ± 8.64 \\ \hline
\multirow{2}{*}{\begin{tabular}[c]{@{}c@{}}20 vCPUs\\ 1 GiB\end{tabular}} & Seq. & 2369.20 ± 245.57 & 4277.00 ± 119.25 & 8545.20 ± 133.26 \\ \cline{2-5} 
 & Para. & 404.00 ± 31.21 & 330.20 ± 3.03 & 1458.40 ± 161.15 \\ \hline
\multirow{2}{*}{\begin{tabular}[c]{@{}c@{}}20 vCPUs\\ 2 GiB\end{tabular}} & Seq. & 4587.20 ± 751.16 & 8615.60 ± 198.47 & 17970.60 ± 1701.12 \\ \cline{2-5} 
 & Para. & 764.40 ± 18.47 & 679.40 ± 72.50 & 2888.80 ± 42.47 \\ \hline
\end{tabular}
\end{table}
}

To assess how parallelization speeds up different stages in \vc (free page allocation, page color identification, and insertion into colored free page lists), we compare its parallel and sequential versions in a local VM with 20 vCPUs pinned to an LLC domain. We measure construction times for 500 MiB, 1 GiB, and 2 GiB lists. Table~\ref{tab: vcolorev} shows that parallelization reduces total times to 1.02 s, 2.19 s, and 4.33 s, delivering speedups of $6.4\times$, $6.7\times$, and $7.1\times$ over the sequential version. These short runtimes make regular list refills feasible with low overhead in popular VM sizes. Accuracy is verified via the custom hypercall exposing GPA-to-HPA mappings, achieving 100\% correct color identification. Last, we measure page allocation latency by using fio to trigger page cache misses. Averaged over 100,000 samples, the latency is 1,761 ns, lower than the 1,887 ns of the default page allocator.

\subsection{Performance and Accuracy of \vs}
\label{sec: vsetev}

\subsubsection*{LLC Sets Coverage}
\vs aims to monitor one set for each LLC \textit{row}. On our Skylake-SP CPU, for example, the uncontrollable HPA bit 16 causes the offset partition in each color group to produce an eviction set in one of two LLC rows (Figure~\ref{fig:vsetparabuild}). To cover both rows, \vs must increase $f$ (the number of eviction sets built per offset partition per color group). The theoretical coverage is $100\%\!\times\!(1\!-\!P_f)\!+\!50\%\!\times\!P_f$, where $P_f$, the probability of covering one of the two rows with $f$, is $\smash{C_{n}^{f}} / \smash{C_{2n}^{f}}$, and $n$ is the number of slices. To measure actual coverage, we run experiments in a local 10-vCPU VM pinned to an LLC domain, varying $f$. Coverage is inferred via the GPA-to-HPA hypercall, and \vs overhead is measured from both eviction set construction and monitoring (prime and probe phases), as shown in Table~\ref{tab: vsettargetset}. The measured coverage matches the theoretical one, with diminishing gains as $f$ increases while overhead continues to grow. We choose $f\!=\!4$ as the optimal trade-off between coverage and overhead.

{
\renewcommand{\arraystretch}{1.1}
\begin{table}[h!]
\setlength{\tabcolsep}{2pt}
\centering
\footnotesize
\caption{\small LLC Coverage and Overhead vs. $f$ in \vs.}
\label{tab: vsettargetset}
\begin{tabular}{|c|c|c|c|c|c|}
\hline
\textbf{Metrics} & $\boldsymbol{f\!=\!2}$ & $\boldsymbol{f\!=\!3}$ & $\boldsymbol{f\!=\!4}$ & $\boldsymbol{f\!=\!5}$ & $\boldsymbol{f\!=\!6}$ \\ \hline
Theo. Cov. & 75.64\% & 88.46\% & 94.70\% & 97.64\% & 98.99\% \\ \hline
Exp. Cov. & 75.94\% & 88.75\% & 94.69\% & 96.88\% & 99.69\%  \\ \hline
Build (s) & 1.06 ± 0.29 & 1.93 ± 0.47 & 2.02 ± 0.28 & 2.58 ± 0.41 & 3.39 ± 0.68 \\ \hline
Monitor (ms) & 1.31 ± 0.12 & 1.99 ± 0.22 & 2.92 ± 0.21 & 3.60 ± 0.28 & 4.31 ± 0.29 \\ \hline
\end{tabular}
\end{table}
}

\subsubsection*{Prime+Probe Overhead}

To evaluate overhead and parallelization benefits of periodic \textit{windowed Prime+Probe}, we measure prime and probe times as \vs monitors 4,096 LLC sets ($f\!=\!4$) in local VMs of varying sizes. With 2 vCPUs, Prime+Probe runs sequentially: one thread pair primes all sets, waits, then probes them. With 10 vCPUs, five pairs each handle 20\% of sets in parallel; with 20 vCPUs, ten pairs each handle 10\%. Table~\ref{tab: ppoverhead} shows prime and probe times drop nearly linearly as vCPUs increase. The wait window accounts for prime time; for example, with 10 vCPUs and a 7 ms window, probing starts ~5.62 ms after the last set primes, ensuring the first set is probed after 7 ms. Since prime–probe time differences are under 1 ms, delays for the last sets are negligible. The full cycle (prime+wait+probe) is under 10 ms, keeping monitoring overhead below 1\% at a 1s interval. Only small VMs risk violating the wait window if prime time exceeds it.

{
\renewcommand{\arraystretch}{0.3}
\captionsetup{belowskip=-4pt}
\begin{table}[h!]
    \setlength{\tabcolsep}{2pt}
    \footnotesize
    \centering
    \caption{\small Prime+Probe Overhead.}
    \begin{tabular}[t]{lcc}
    \toprule \textbf{Settings} & \textbf{Prime (ms)} & \textbf{Probe (ms)} \\
    \midrule
    Sequential, 2 vCPUs & 7.23 ± 0.65 & 7.79 ± 0.35 \\
    \midrule
    Parallel, 10 vCPUs & 1.38 ± 0.22 & 1.61 ± 0.10 \\
    \midrule
    Parallel, 20 vCPUs & 0.65 ± 0.08 & 0.82 ± 0.02 \\
    \bottomrule
    \end{tabular}
    \label{tab: ppoverhead}
\end{table}%
}

\subsubsection*{Prime+Probe Accuracy}
To assess probe phase accuracy and how EWMA smooths results, we run \vs with a 1s interval in a local 10-vCPU VM pinned to 10 cores in an LLC domain across three phases: \textit{manual}, \textit{idle}, and \textit{nginx}. During the manual phase, \vs uses \textit{windowless Prime+Probe} to monitor one LLC set; before probing, 2, 4, 6, 8, and 11 cache lines are flushed from that set over time. Figure~\ref{fig:evrateandwindow}a shows accurate detection of eviction percentages. Then \vs switches back to \textit{windowed Prime+Probe} ($f\!=\!4$). In the idle phase, with the host idle, \vs with a 7 ms window manages to detect 1–2 evictions per set per interval. In the last phase, nginx runs in a co-located VM pinned to the other 10 cores in the same LLC domain, and EWMA values respond promptly to changes while effectively smoothing spikes.

\subsubsection*{Wait Window Sensitivity}
\begin{figure}[t]
    \centering
    \subfloat[\centering EWMA eviction percentages]{{\includegraphics[width=0.449\columnwidth]{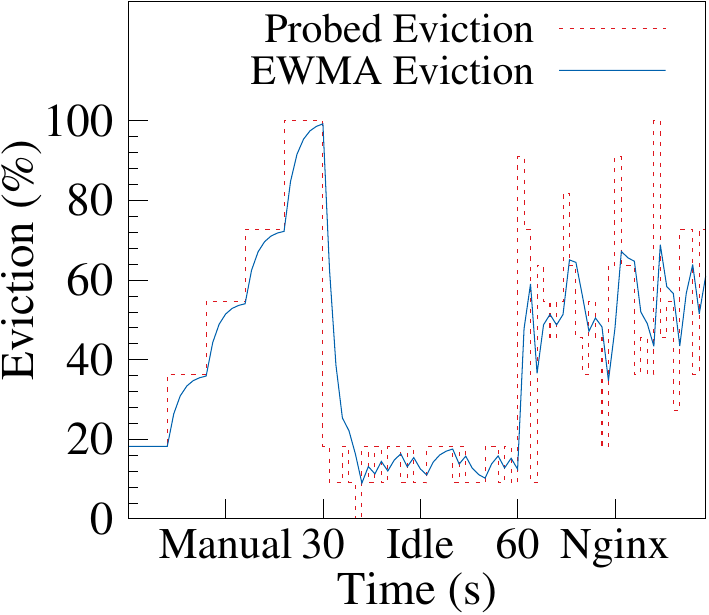} }}%
    \qquad
    \subfloat[\centering Evictions vs. Wait window]{{\includegraphics[width=0.441\columnwidth]{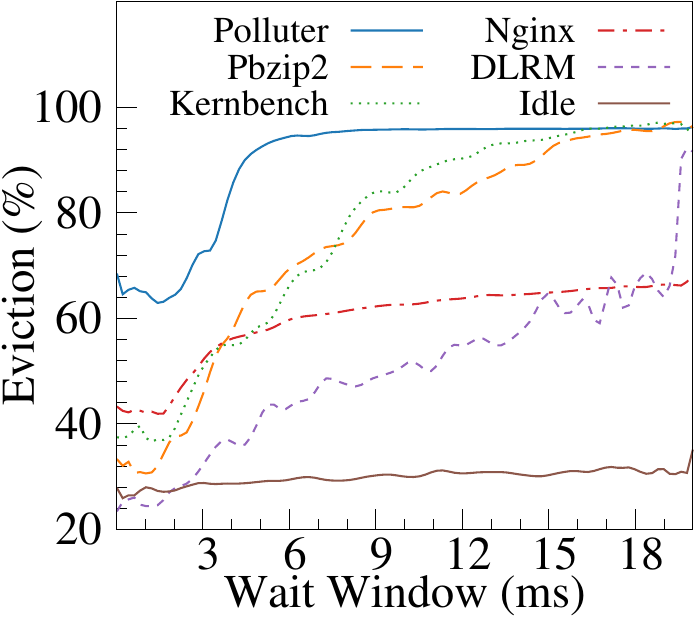} }}%
    \caption{\vs's accuracy and sensitivity to wait windows.}%
    \label{fig:evrateandwindow}%
\end{figure}

We set the default wait window to 7 ms after evaluating how probed eviction rates vary with window size. We create two local 10-vCPU VMs pinned to 20 cores within an LLC domain: one runs \vs, the other runs workloads that generate heavy contention (\textit{cache polluter}), moderate contention (pbzip2, kernbench), and light contention (nginx, DLRM). Figure~\ref{fig:evrateandwindow}b shows the results. Since the average prime time is 1.38 ms, the waits with 0-1.38 ms windows are effectively the same. Eviction rates probed with windows longer than 7 ms underestimate contention, as sets may already be fully evicted before probing. However, shorter windows lack resolution. For instance, 0-3 ms windows fail to distinguish idle from moderate contention in DLRM: the probed eviction percentage rises with window size for DLRM but stays flat with an idle neighbor, where background tasks may reuse the same cache lines over time. Notably, while vCPU preemptions can prolong the wait, any resulting full evictions trigger window shrinking, which in turn reduces mid-monitoring preemptions but may lower resolution, especially for VMs with frequent preemptions.

\subsection{LLC Contention and Page Colors in Clouds}
\label{sec:cloudprobe}

\begin{figure}[t] 
\centering \subfloat[\centering Dynamic LLC contention]{{\includegraphics[width=0.445\columnwidth]{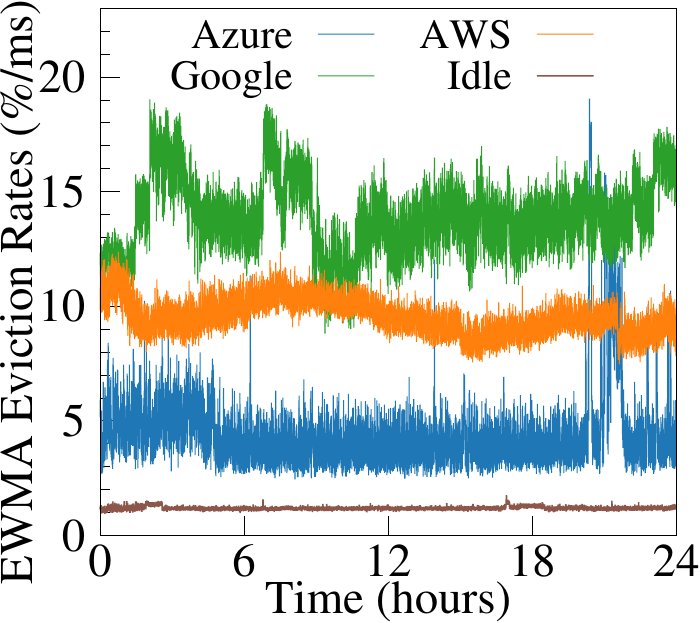} }}%
\qquad \subfloat[\centering Asymmetric LLC contention]{{\includegraphics[width=0.445\columnwidth]{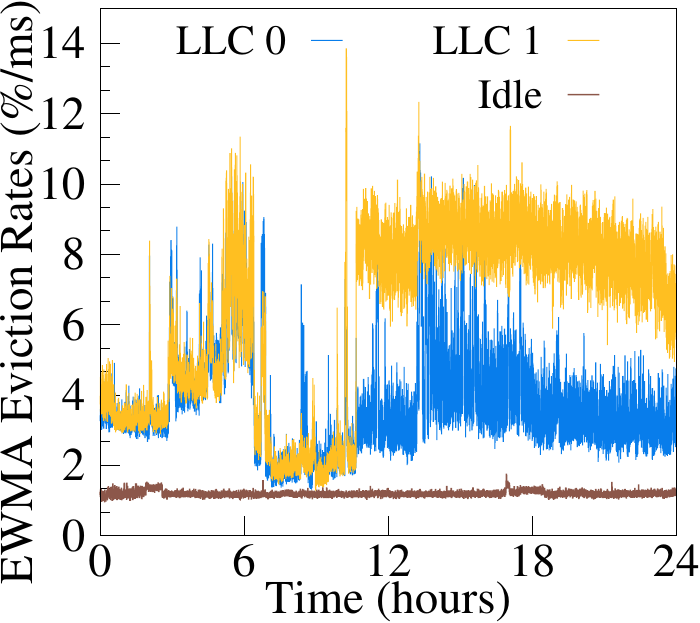} }}%
\caption{\small Dynamic and asymmetric LLC contention in clouds.}%
\label{fig:cloudllcev}%
\end{figure}

\begin{figure}[h!]
    \centering    \includegraphics[width=1\columnwidth]{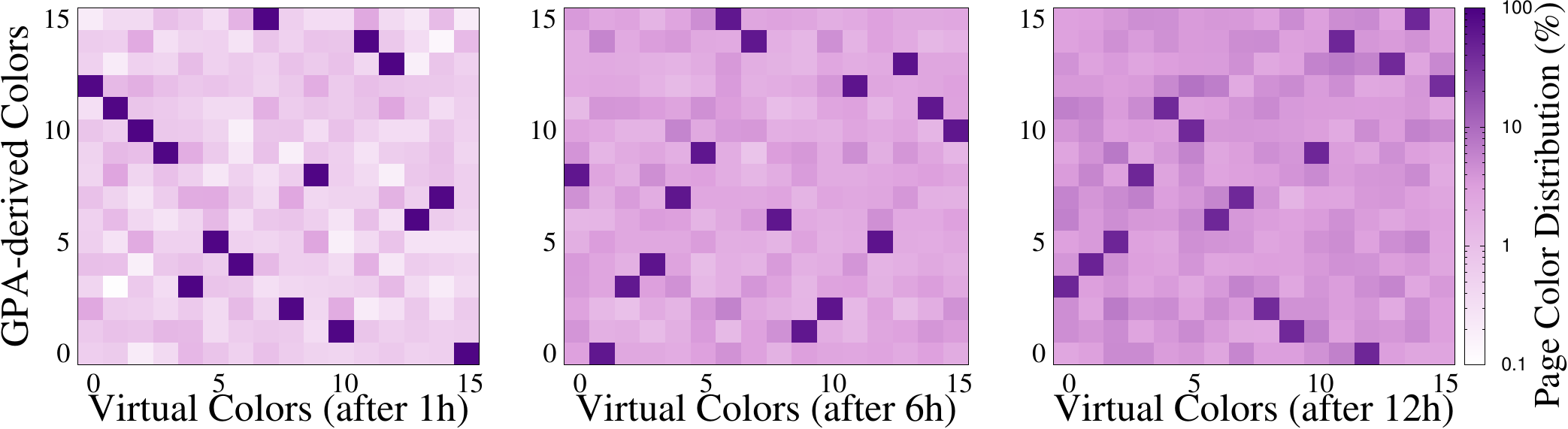}
    \caption{\small Page color distribution skews over time in clouds. We rebuild the color filters before each \vc run and the same virtual color may map to different real L2 colors across runs.}
    \label{fig:cloudcolorev}
\end{figure}

We probe three cloud VMs with \vs (1s interval, 24 h) to observe dynamic LLC contention. Figure \ref{fig:cloudllcev}a shows that AWS and Google VMs consistently observe higher eviction rates than those observed by an 8-vCPU local VM on an idle host, indicating persistent contention. In contrast, the Azure VM appears to run on a mostly quiescent host, with only a brief contention spike toward the end. We suspect that the higher contention on the Google VM stems from its placement on a 28-core CPU, which can host more contending VMs on the other 20 cores, and we observe the wait window shrink by 1 ms twice in the first 12 hours, triggered by full evictions across sets. In contrast, the AWS and Azure VMs run on 26-core CPUs. To demonstrate asymmetric LLC contention in clouds, we repeat the experiment on a Google VM with 40 vCPUs evenly distributed across two LLC domains on two 28-core CPUs. As shown in Figure~\ref{fig:cloudllcev}b, the domains exhibit nearly identical contention for the first 10.5 hours, followed by 13.5 hours during which LLC 1 experiences higher contention than LLC 0. To assess page color reliability, we probe the 8-vCPU Google VM with \vc, which filters 21,120 pages every 15 minutes, interleaved with repeated kernbench runs. Initially, 100\% of pages sharing the same GPA-derived color also share the same virtual color. After one hour, the overlap drops to an average of 89\%, and after 12 hours it falls to 43\% (Figure \ref{fig:cloudcolorev}). This skew appears region-specific: the same Google VM, when aged in the \texttt{us-central} region, maintains stable page colors, suggesting region-dependent policies. These results indicate that guest physical pages are occasionally remapped, which can break eviction sets and invalidate virtual colors. To remain correct, \vc should rebuild its color filters and free page lists, and \vs should reconstruct LLC eviction sets at least once per hour, a feasible strategy given their previously demonstrated low cost.

\subsection{Reduced Inter-VM Interference with \lcas}
\label{sec: lcasev}

\begin{figure}[h!]
    \centering
    \includegraphics[width=1\columnwidth]{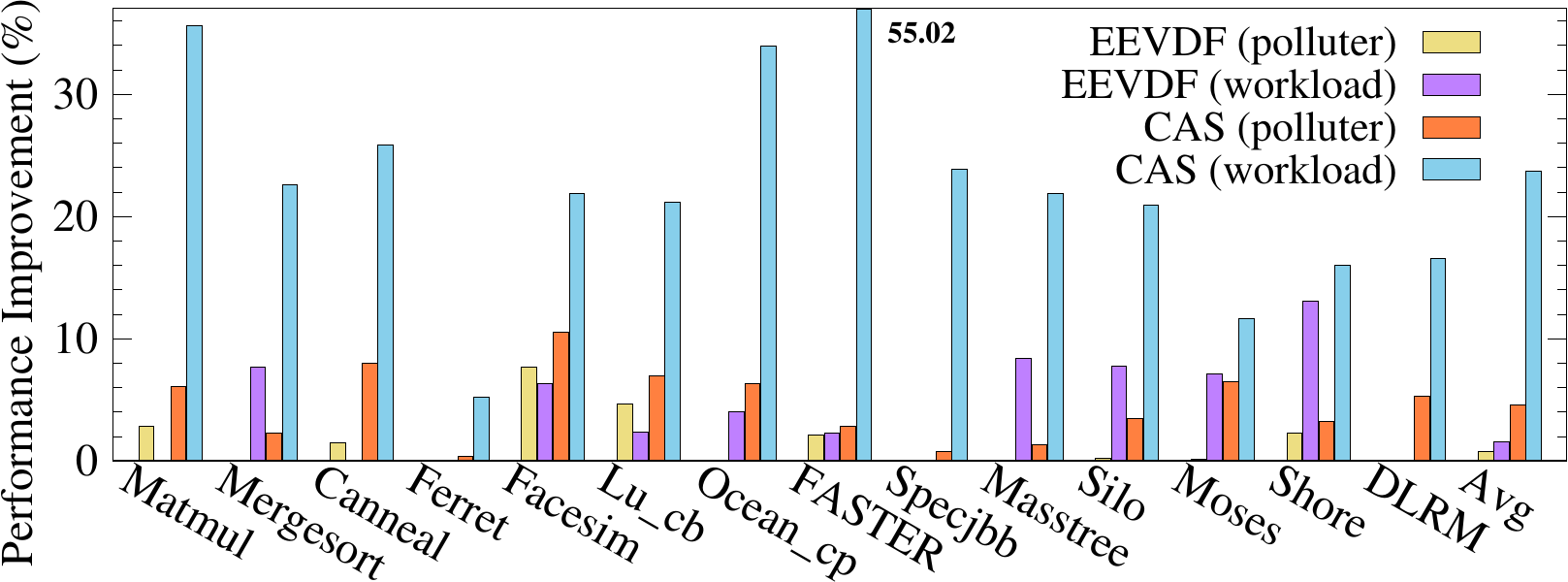}
    \caption{\small Performance improvement compared to scx\_rusty.}
    \label{fig:ivhtest}
\end{figure}

We evaluate \lcas on a 16-vCPU local VM, split evenly across two LLC domains with 8 vCPUs pinned to 8 cores in each domain. To introduce asymmetric contention, we create two additional 12-vCPU VMs: one pinned to the remaining 12 cores in the first LLC domain, running the cache polluter, and the other pinned to the remaining 12 cores in the second domain, running sysbench with minimal LLC accesses. Experiments use a suite of cache-sensitive workloads under three schedulers: scx\_rusty, EEVDF, and \lcas, with each workload limited to 8 threads to demonstrate the impact of LLC-aware task placement in an underloaded system. Since scx\_rusty currently targets multi-CCX systems~\cite{amdccx}, we enable NUMA interleave to simulate two LLCs in one NUMA node, ensuring fair comparison with EEVDF. As shown in Figure~\ref{fig:ivhtest}, on average, EEVDF performs comparably to \texttt{scx\_rusty}, while \lcas improves performance by 24.8\%. Notably, the cache polluter achieves 4.5\% higher throughput with \lcas, demonstrating that neighbor VMs also benefit from reduced inter-VM cache conflicts. Moreover, workload performance fluctuates under EEVDF and scx\_rusty due to occasional placement on either domain, whereas \lcas consistently prioritizes the less-contended domain. For example, \lcas schedules Silo only 16\% of the time on the polluted domain, compared to 40-60\% for EEVDF and scx\_rusty.

\subsection{Reduced Intra-VM Interference with \vcpc}
\label{sec: vcpcev}

\begin{figure}[h!]
    \centering
    \includegraphics[width=1\columnwidth]{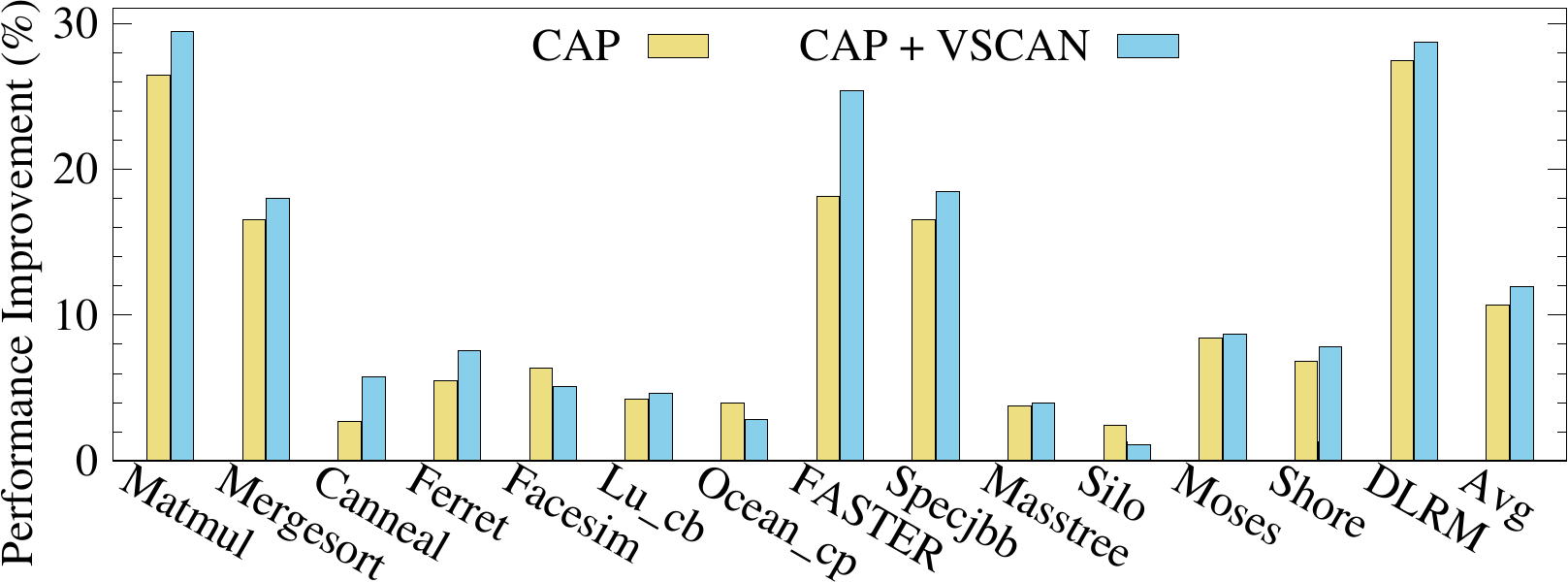}
    \caption{\small Performance improvement compared to vanilla Linux.}
    \label{fig:vcpcev}
\end{figure}

We evaluate \vcpc on a local 10-vCPU VM pinned to 10 cores in one LLC domain. Nine vCPUs run cache-sensitive workloads, while one vCPU runs fio scanning an 80 MB file to introduce intra-VM LLC pollution via the page cache. To create asymmetric per-color contention, another 10-vCPU VM is co-located on the remaining 10 cores in the same LLC domain, running the cache poisoner to stress one LLC zone corresponding to a virtual color. Experiments are performed under three settings: vanilla Linux, \vcpc, and \vcpc combined with \vs. As shown in Figure~\ref{fig:vcpcev}, \vcpc improves workload performance by an average of 10.7\% without reducing fio’s throughput. With \vs reporting per-color contention, \vcpc can preferentially allocate pages from the hottest color stressed by the cache poisoner, yielding up to 7.3\% additional improvement with FASTER. Workloads like Silo see a minor performance drop due to \vs's overhead (§~\ref{sec: overhead}). Although the average extra gain is modest (1\%), this shows the potential of intra-VM partitioning in large LLCs to let cache-\textit{insensitive} workloads absorb inter-VM interference. 

\subsection{Overhead of CacheX}
\label{sec: overhead}

\begin{figure}[h!]
    \centering
    \includegraphics[width=1\columnwidth]{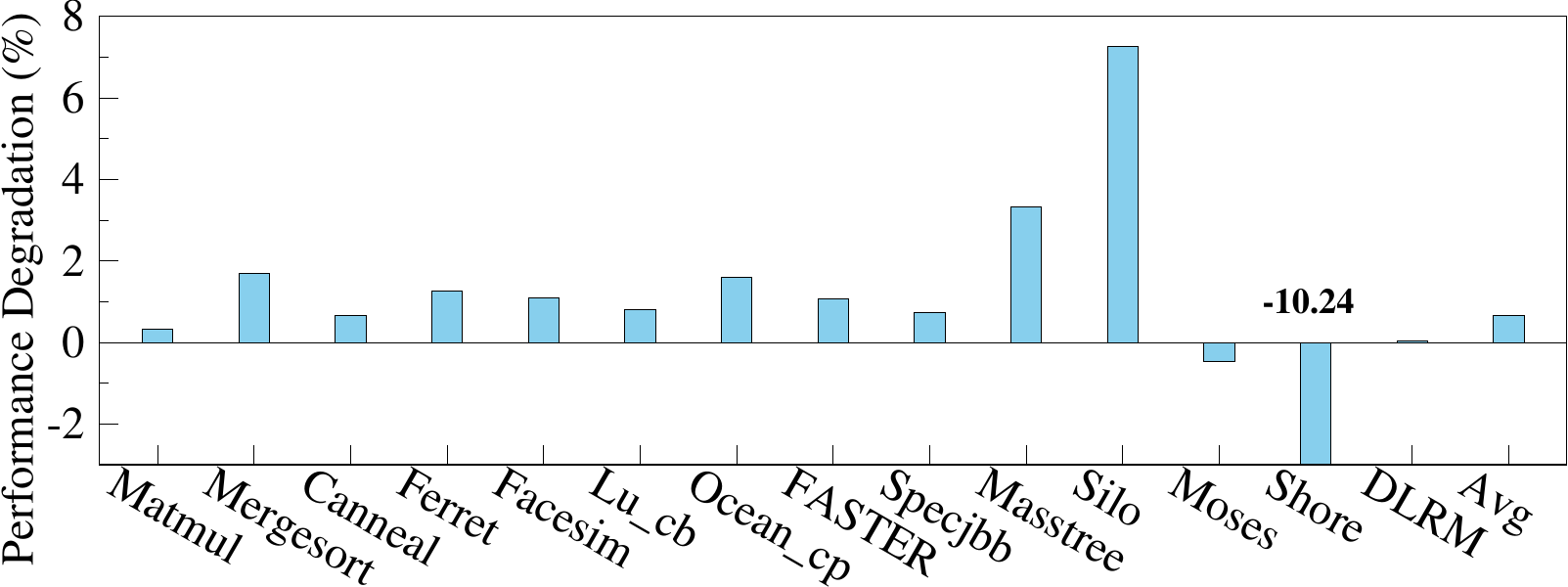}
    \caption{\small Performance degradation refers to throughput reduction or latency increase with \vs.} 
    \label{fig:overhead}
\end{figure}

\ev incurs only the periodic overhead from \vs. To quantify this, we create a 10-vCPU VM pinned to 10 cores in a single LLC domain and run the same workload suite used in \lcas and \vcpc, with and without \vs (1s interval). Since workloads do not benefit from \ev in this setup, any performance difference reflects its overhead. As shown in Figure~\ref{fig:overhead}, \vs introduces minimal overhead, averaging a 0.66\% performance degradation. Throughput-oriented workloads are slightly affected, while latency-critical workloads with small tasks are more sensitive. For example, the p95 latency of Silo may increase due to \vs’s high-priority monitoring, whereas workloads like Shore can benefit from core frequency warm-up before execution.

\section{Related Work}

\noindent \textbf{Hypervisor-assisted optimizations} are proposed to improve both transparency and control of vCache. vLLC~\cite{kim2015vcache} extends TLB entries with GPAs for LLC indexing, enabling guest-side page coloring. CoPlace~\cite{shang2021coplace} aligns guest and host physical page colors by adjusting host-side allocation, avoiding skewed page color distribution in VMs. CacheSlicer~\cite{shahrad2021provisioning} allows users to allocate dedicated LLC ways to cache-sensitive VMs. vCAT~\cite{xu2017vcat} exposes virtualized CAT partitions for intra-VM cache partitioning. CPS~\cite{liu2023cps} exposes vCPU-to-LLC slice mappings to enable NUCA (Non-Uniform Cache Access)-aware scheduling within VMs. However, in the multi-cloud era~\cite{yang2023skypilot}, adopting these approaches is challenging due to the heterogeneity and proprietary nature of hypervisors. Furthermore, granting multiple VMs control over shared caches can result in conflicting optimization goals. \smallskip

\noindent \textbf{Profiling techniques} are developed to infer opaque vCache characteristics without any hypervisor or hardware support. vTop~\cite{vsched} uses cache line transfer latency to detect the dynamic vCPU topology, allowing the task scheduler to group dependent tasks into vCPUs that share L2 cache or LLC. CacheInspector~\cite{song2021cacheinspector} estimates the dynamic LLC size in cloud VMs by searching for a buffer size that, when accessed, hits halfway of the throughput cliff between LLC and memory. However, it cannot report fine-grained information like per-LLC size for system-level optimizations, and its accuracy struggles with misaligned huge pages~\cite{fakehuge} and changing vCPU topology~\cite{elbing2020linux}, both of which are common in clouds.  \smallskip

\noindent \textbf{Eviction sets} are primarily used in cache-based side-channel attacks~\cite{irazoqui2015s, liu2015last, oren2015spy, inci2016cache, gras2017aslr, yan2019attack, kocher2020spectre} and their defenses~\cite{volos2024principled, hofmann2024gaussian}. Many recent efforts focus on accelerating their construction~\cite{vila2019theory, purnal2021prime+, zhao2024last, kessous2024prune+, morgan2025slice+}. We are the first to apply eviction sets in cache probing and to parallelize their construction in cloud VMs. A major countermeasure against cache-based side channel attacks involves redesigning cache architectures, such as randomized caches~\cite{song2021randomized, qureshi2018ceaser}, skewed randomized caches~\cite{werner2019scattercache, qureshi2019new}, or fully associative caches~\cite{saileshwar2021mirage}. These designs may make any eviction set construction, including ours, significantly more difficult or even infeasible. However, they require fundamental hardware changes unlikely to see near-term deployment in public clouds.

\section{Conclusion}

This paper introduces novel probing techniques that uncover accurate, fine-grained cache characteristics and page colors, and reveal their dynamic nature in cloud VMs. The probed information enhances cache-based optimizations, improving workload throughput and latency in dynamic multi-cloud environments. We plan to extend these techniques to more processors for broader cloud adoption.

\bibliographystyle{ACM-Reference-Format}
\bibliography{references}

@inproceedings{gras2017aslr,
  series = {NDSS 2017},
  title = {ASLR on the Line: Practical Cache Attacks on the MMU},
  url = {http://dx.doi.org/10.14722/ndss.2017.23271},
  DOI = {10.14722/ndss.2017.23271},
  booktitle = {Proceedings 2017 Network and Distributed System Security Symposium},
  publisher = {Internet Society},
  author = {Gras,  Ben and Razavi,  Kaveh and Bosman,  Erik and Bos,  Herbert and Giuffrida,  Cristiano},
  year = {2017},
  collection = {NDSS 2017}
}

@inbook{inci2016cache,
  title = {Cache Attacks Enable Bulk Key Recovery on the Cloud},
  ISBN = {9783662531402},
  ISSN = {1611-3349},
  url = {http://dx.doi.org/10.1007/978-3-662-53140-2_18},
  DOI = {10.1007/978-3-662-53140-2_18},
  booktitle = {Cryptographic Hardware and Embedded Systems – CHES 2016},
  publisher = {Springer Berlin Heidelberg},
  author = {İnci,  Mehmet Sinan and Gulmezoglu,  Berk and Irazoqui,  Gorka and Eisenbarth,  Thomas and Sunar,  Berk},
  year = {2016},
  pages = {368–388}
}

@inproceedings{irazoqui2015s,
  title = {S\$A: A Shared Cache Attack That Works across Cores and Defies VM Sandboxing -- and Its Application to AES},
  url = {http://dx.doi.org/10.1109/SP.2015.42},
  DOI = {10.1109/sp.2015.42},
  booktitle = {2015 IEEE Symposium on Security and Privacy},
  publisher = {IEEE},
  author = {Irazoqui,  Gorka and Eisenbarth,  Thomas and Sunar,  Berk},
  year = {2015},
  month = may,
  pages = {591–604}
}

@inproceedings{bui2019extended,
  series = {EuroSys ’19},
  title = {When eXtended Para - Virtualization (XPV) Meets NUMA},
  url = {http://dx.doi.org/10.1145/3302424.3303960},
  DOI = {10.1145/3302424.3303960},
  booktitle = {Proceedings of the Fourteenth EuroSys Conference 2019},
  publisher = {ACM},
  author = {Bui,  Bao and Mvondo,  Djob and Teabe,  Boris and Jiokeng,  Kevin and Wapet,  Lavoisier and Tchana,  Alain and Thomas,  Gaël and Hagimont,  Daniel and Muller,  Gilles and DePalma,  Noel},
  year = {2019},
  month = mar,
  pages = {1–15},
  collection = {EuroSys ’19}
}

@inproceedings{zhang2009towards,
  series = {EuroSys ’09},
  title = {Towards Practical Page Coloring-based Multicore Cache Management},
  url = {http://dx.doi.org/10.1145/1519065.1519076},
  DOI = {10.1145/1519065.1519076},
  booktitle = {Proceedings of the 4th ACM European conference on Computer systems},
  publisher = {ACM},
  author = {Zhang,  Xiao and Dwarkadas,  Sandhya and Shen,  Kai},
  year = {2009},
  month = apr,
  pages = {89–102},
  collection = {EuroSys ’09}
}

@inproceedings{shang2021coplace,
  title = {CoPlace: Effectively Mitigating Cache Conflicts in Modern Clouds},
  url = {http://dx.doi.org/10.1109/PACT52795.2021.00027},
  DOI = {10.1109/pact52795.2021.00027},
  booktitle = {2021 30th International Conference on Parallel Architectures and Compilation Techniques (PACT)},
  publisher = {IEEE},
  author = {Shang,  Xiaowei and Jia,  Weiwei and Shan,  Jianchen and Ding,  Xiaoning},
  year = {2021},
  month = sep,
  pages = {274–288}
}

@inproceedings{song2021randomized,
  title = {Randomized Last-Level Caches Are Still Vulnerable to Cache Side-Channel Attacks! But We Can Fix It},
  url = {http://dx.doi.org/10.1109/SP40001.2021.00050},
  DOI = {10.1109/sp40001.2021.00050},
  booktitle = {2021 IEEE Symposium on Security and Privacy (SP)},
  publisher = {IEEE},
  author = {Song,  Wei and Li,  Boya and Xue,  Zihan and Li,  Zhenzhen and Wang,  Wenhao and Liu,  Peng},
  year = {2021},
  month = may,
  pages = {955–969}
}

@inproceedings{qureshi2018ceaser,
  title = {CEASER: Mitigating Conflict-Based Cache Attacks via Encrypted-Address and Remapping},
  url = {http://dx.doi.org/10.1109/MICRO.2018.00068},
  DOI = {10.1109/micro.2018.00068},
  booktitle = {2018 51st Annual IEEE/ACM International Symposium on Microarchitecture (MICRO)},
  publisher = {IEEE},
  author = {Qureshi,  Moinuddin K.},
  year = {2018},
  month = oct,
  pages = {775–787}
}

@misc{mccalpin2021mapping,
  doi = {10.26153/TSW/14539},
  url = {https://repositories.lib.utexas.edu/handle/2152/87595},
  author = {McCalpin,  John D.},
  language = {en},
  title = {Mapping Addresses to L3/CHA Slices in Intel Processors},
  publisher = {The University of Texas at Austin},
  year = {2021},
  copyright = {Attribution-ShareAlike 3.0 United States}
}

@inproceedings{soori2024nucalloc,
  series = {ICS ’24},
  title = {NUCAlloc: Fine-Grained Block Placement in Hashed Last-Level NUCA Caches},
  url = {http://dx.doi.org/10.1145/3650200.3656604},
  DOI = {10.1145/3650200.3656604},
  booktitle = {Proceedings of the 38th ACM International Conference on Supercomputing},
  publisher = {ACM},
  author = {Soori,  Raveendra and Prabhu,  Shreyas and Chawla,  Harpreet Singh and Ferdman,  Michael},
  year = {2024},
  month = may,
  pages = {85–97},
  collection = {ICS ’24}
}

@inproceedings {werner2019scattercache,
author = {Mario Werner and Thomas Unterluggauer and Lukas Giner and Michael Schwarz and Daniel Gruss and Stefan Mangard},
title = {{ScatterCache}: Thwarting Cache Attacks via Cache Set Randomization},
booktitle = {28th USENIX Security Symposium (USENIX Security 19)},
year = {2019},
isbn = {978-1-939133-06-9},
address = {Santa Clara, CA},
pages = {675--692},
url = {https://www.usenix.org/conference/usenixsecurity19/presentation/werner},
publisher = {USENIX Association},
month = aug
}

@inproceedings{qureshi2019new,
  series = {ISCA ’19},
  title = {New attacks and defense for encrypted-address cache},
  url = {http://dx.doi.org/10.1145/3307650.3322246},
  DOI = {10.1145/3307650.3322246},
  booktitle = {Proceedings of the 46th International Symposium on Computer Architecture},
  publisher = {ACM},
  author = {Qureshi,  Moinuddin K.},
  year = {2019},
  month = jun,
  pages = {360–371},
  collection = {ISCA ’19}
}

@inproceedings{purnal2021prime+,
  series = {CCS ’21},
  title = {Prime+Scope: Overcoming the Observer Effect for High-Precision Cache Contention Attacks},
  url = {http://dx.doi.org/10.1145/3460120.3484816},
  DOI = {10.1145/3460120.3484816},
  booktitle = {Proceedings of the 2021 ACM SIGSAC Conference on Computer and Communications Security},
  publisher = {ACM},
  author = {Purnal,  Antoon and Turan,  Furkan and Verbauwhede,  Ingrid},
  year = {2021},
  month = nov,
  pages = {2906–2920},
  collection = {CCS ’21}
}

@inproceedings{liu2015last,
  title = {Last-Level Cache Side-Channel Attacks are Practical},
  url = {http://dx.doi.org/10.1109/SP.2015.43},
  DOI = {10.1109/sp.2015.43},
  booktitle = {2015 IEEE Symposium on Security and Privacy},
  publisher = {IEEE},
  author = {Liu,  Fangfei and Yarom,  Yuval and Ge,  Qian and Heiser,  Gernot and Lee,  Ruby B.},
  year = {2015},
  month = may,
  pages = {605–622}
}

@inproceedings{fakehuge,
  series = {EuroSys ’23},
  title = {Making Dynamic Page Coalescing Effective on Virtualized Clouds},
  url = {http://dx.doi.org/10.1145/3552326.3567487},
  DOI = {10.1145/3552326.3567487},
  booktitle = {Proceedings of the Eighteenth European Conference on Computer Systems},
  publisher = {ACM},
  author = {Jia,  Weiwei and Zhang,  Jiyuan and Shan,  Jianchen and Ding,  Xiaoning},
  year = {2023},
  month = may,
  pages = {298–313},
  collection = {EuroSys ’23}
}

@article{song2021cacheinspector,
  title = {CacheInspector: Reverse Engineering Cache Resources in Public Clouds},
  volume = {18},
  ISSN = {1544-3973},
  url = {http://dx.doi.org/10.1145/3457373},
  DOI = {10.1145/3457373},
  number = {3},
  journal = {ACM Transactions on Architecture and Code Optimization},
  publisher = {Association for Computing Machinery (ACM)},
  author = {Song,  Weijia and Delimitrou,  Christina and Shen,  Zhiming and Renesse,  Robbert Van and Weatherspoon,  Hakim and Benmohamed,  Lotfi and Vaulx,  Frederic De and Mahmoudi,  Charif},
  year = {2021},
  month = jun,
  pages = {1–25}
}

@inproceedings{merrifield2016performance,
  series = {VEE ’16},
  title = {Performance Implications of Extended Page Tables on Virtualized x86 Processors},
  url = {http://dx.doi.org/10.1145/2892242.2892258},
  DOI = {10.1145/2892242.2892258},
  booktitle = {Proceedings of the12th ACM SIGPLAN/SIGOPS International Conference on Virtual Execution Environments},
  publisher = {ACM},
  author = {Merrifield,  Timothy and Taheri,  H. Reza},
  year = {2016},
  month = mar,
  pages = {25–35},
  collection = {VEE ’16}
}

@MISC{intelcat,
    author = {{Intel}},
    title = {{Intel (R) RDT Software Package}},
    url  = "https://github.com/intel/intel-cmt-cat",
    date = {2025-08-01},
}

@MISC{amdccx,
    author = {{AMD}},
    title = {{AMD EPYC 9005 Processor Architecture Overview}},
    url  = "https://shorturl.at/mC1rS",
    date = {2025-08-01},
}

@MISC{fio,
    author = {{Jens Axboe}},
    title = {{fio - Flexible I/O tester}},
    url  = "https://fio.readthedocs.io/en/latest/index.html",
    date = {2025-08-01},
}

@MISC{guesttsc,
    author = {{Intel}},
    title = {{Timestamp-Counter Scaling for Virtualization White Paper}},
    url  = "https://kib.kiev.ua/x86docs/Intel/WhitePapers/333159-001.pdf",
    date = {2025-08-01},
}

@MISC{rusty,
    author = {{sched\_ext Docs}},
    title = {{scx\_rusty}},
    url  = "https://sched-ext.com/docs/scheds/rust/scx_rusty",
    date = {2025-08-01},
}

@MISC{schedcache,
    author = {{Chen Yu}},
    title = {{[RFC PATCH 0/5] sched: introduce cache aware scheduling}},
    url  = "https://lkml.org/lkml/2025/4/21/78",
    date = {2025-08-01},
}

@MISC{vPMU,
    author = {{Xiong Zhang}},
    title = {{Introduce Passthrough vPMU}},
    url  = "https://lore.kernel.org/all/20240126085444.324918-1-xiong.y.zhang@linux.intel.com/",
    date = {2025-08-01},
}

@MISC{tscscaling,
    author = {{Intel}},
    title = {{Intel® 64 and IA-32 Architectures Software Developer’s Manual Volume 3C: System Programming Guide, Part 3}},
    url  = "https://www.intel.com/content/dam/www/public/us/en/documents/manuals/64-ia-32-architectures-software-developer-vol-3c-part-3-manual.pdf",
    date = {2025-08-01},
}

@MISC{dlrmcode,
    author = {Facebook},
    title = {{Deep Learning Recommendation Model for Personalization and Recommendation Systems}},
    url  = "https://github.com/facebookresearch/dlrm",
    date = {2025-08-01},
}

@MISC{googlfaas,
    author = {{Google Cloud}},
    title = {{Cloud Run functions}},
    url  = "https://cloud.google.com/functions",
    date = {2025-08-01},
}

@MISC{googlevm,
    author = {Google Cloud},
    title = {{N1 machine series}},
    url  = "https://cloud.google.com/compute/docs/general-purpose-machines#n1_machines",
    date = {2025-08-01},
}

@MISC{azurevm,
    author = {Microsoft Azure},
    title = {{Azure Virtual Machines}},
    url  = "https://learn.microsoft.com/en-us/azure/virtual-machines/sizes/memory-optimized/dv2-dsv2-series-memory",
    date = {2025-08-01},
}

@MISC{awsvm,
    author = {Amazon Web Services (AWS)},
    title = {{Amazon EC2 M5 Instances}},
    url  = "https://aws.amazon.com/ec2/instance-types/m5/",
    date = {2025-08-01},
}

@inproceedings{vsched,
  series = {EuroSys ’25},
  title = {Optimizing Task Scheduling in Cloud VMs with Accurate vCPU Abstraction},
  url = {http://dx.doi.org/10.1145/3689031.3696092},
  DOI = {10.1145/3689031.3696092},
  booktitle = {Proceedings of the Twentieth European Conference on Computer Systems},
  publisher = {ACM},
  author = {Guo,  Edward and Jia,  Weiwei and Ding,  Xiaoning and Shan,  Jianchen},
  year = {2025},
  month = mar,
  pages = {753–768},
  collection = {EuroSys ’25}
}

@MISC{vpmuuncore,
    author = {{Intel}},
    title = {{Profile KVM Kernel and User Space on the KVM System}},
    url  = "https://www.intel.com/content/www/us/en/docs/vtune-profiler/user-guide/2023-0/on-kvm-project-system.html",
    date = {2025-08-01},
}

@MISC{gfpflags,
    author = {{Linux Kernel Documentation}},
    title = {{Memory Allocation Guide}},
    url  = "https://www.kernel.org/doc/html/next/core-api/memory-allocation.html",
    date = {2025-08-01},
}

@MISC{bpfmap,
    author = {{Linux Kernel Documentation}},
    title = {{BPF maps}},
    url  = "https://docs.kernel.org/bpf/maps.html",
    date = {2025-08-01},
}

@inproceedings{soares2008reducing,
  title = {Reducing the Harmful Effects of Last-Level Cache Polluters with an OS-level,  Software-Only Pollute Buffer},
  url = {http://dx.doi.org/10.1109/MICRO.2008.4771796},
  DOI = {10.1109/micro.2008.4771796},
  booktitle = {2008 41st IEEE/ACM International Symposium on Microarchitecture},
  publisher = {IEEE},
  author = {Soares,  Livio and Tam,  David and Stumm,  Michael},
  year = {2008},
  month = nov,
  pages = {258–269}
}

@inproceedings{ding2011srm,
  series = {EuroSys ’11},
  title = {SRM-buffer: An OS Buffer Management Technique to Prevent Last Level Cache from Thrashing in Multicores},
  url = {http://dx.doi.org/10.1145/1966445.1966468},
  DOI = {10.1145/1966445.1966468},
  booktitle = {Proceedings of the sixth conference on Computer systems},
  publisher = {ACM},
  author = {Ding,  Xiaoning and Wang,  Kaibo and Zhang,  Xiaodong},
  year = {2011},
  month = apr,
  pages = {243–256},
  collection = {EuroSys ’11}
}

@MISC{glibcmemcpythreshold,
    author = {{Wangshuo}},
    title = {{Analysis of the VM Performance Deterioration When Running memcpy to Copy 1,000 Bytes in the x86\_64 Environment}},
    url  = "http://bit.ly/3X7Dp8d",
    date = {2021-01-13},
}

@misc{memcpycachesize,
  doi = {10.5446/46877},
  url = {https://av.tib.eu/media/46877},
  author = {Faggioli,  Dario},
  language = {en},
  title = {Reaching "EPYC" Virtualization Performance: Case Study: Tuning VMs for Best Performance on AMD EPYC 7002/7004 Processor Series Based Servers},
  publisher = {FOSDEM VZW},
  year = {2020}
}

@inproceedings{kim2015vcache,
  series = {MICRO-48},
  title = {vCache: Architectural Support for Transparent and Isolated Virtual LLCs in Virtualized Environments},
  url = {http://dx.doi.org/10.1145/2830772.2830825},
  DOI = {10.1145/2830772.2830825},
  booktitle = {Proceedings of the 48th International Symposium on Microarchitecture},
  publisher = {ACM},
  author = {Kim,  Daehoon and Kim,  Hwanju and Kim,  Nam Sung and Huh,  Jaehyuk},
  year = {2015},
  month = dec,
  pages = {623–634},
  collection = {MICRO-48}
}

@article{kessler1992page,
  title = {Page Placement Algorithms for Large Real-Indexed Caches},
  volume = {10},
  ISSN = {1557-7333},
  url = {http://dx.doi.org/10.1145/138873.138876},
  DOI = {10.1145/138873.138876},
  number = {4},
  journal = {ACM Transactions on Computer Systems},
  publisher = {Association for Computing Machinery (ACM)},
  author = {Kessler,  R. E. and Hill,  Mark D.},
  year = {1992},
  month = nov,
  pages = {338–359}
}

@inproceedings {zhong2024managing,
author = {Yuhong Zhong and Daniel S. Berger and Carl Waldspurger and Ryan Wee and Ishwar Agarwal and Rajat Agarwal and Frank Hady and Karthik Kumar and Mark D. Hill and Mosharaf Chowdhury and Asaf Cidon},
title = {Managing Memory Tiers with {CXL} in Virtualized Environments},
booktitle = {18th USENIX Symposium on Operating Systems Design and Implementation (OSDI 24)},
year = {2024},
isbn = {978-1-939133-40-3},
address = {Santa Clara, CA},
pages = {37--56},
url = {https://www.usenix.org/conference/osdi24/presentation/zhong-yuhong},
publisher = {USENIX Association},
month = jul
}

@inproceedings{xu2017vcat,
  title = {vCAT: Dynamic Cache Management Using CAT Virtualization},
  url = {http://dx.doi.org/10.1109/RTAS.2017.15},
  DOI = {10.1109/rtas.2017.15},
  booktitle = {2017 IEEE Real-Time and Embedded Technology and Applications Symposium (RTAS)},
  publisher = {IEEE},
  author = {Xu,  Meng and Thi,  Linh and Phan,  Xuan and Choi,  Hyon-Young and Lee,  Insup},
  year = {2017},
  month = apr,
  pages = {211–222}
}

@inproceedings{elbing2020linux,
  title = {The Linux Load Balance: Wasted vCPUs in Clouds},
  url = {http://dx.doi.org/10.1109/IEEECloudSummit48914.2020.00035},
  DOI = {10.1109/ieeecloudsummit48914.2020.00035},
  booktitle = {2020 IEEE Cloud Summit},
  publisher = {IEEE},
  author = {Elbing,  Matthew and Shan,  Jianchen},
  year = {2020},
  month = oct,
  pages = {174–175}
}

@inproceedings{lozi2016linux,
  series = {EuroSys ’16},
  title = {The Linux Scheduler: a Decade of Wasted Cores},
  url = {http://dx.doi.org/10.1145/2901318.2901326},
  DOI = {10.1145/2901318.2901326},
  booktitle = {Proceedings of the Eleventh European Conference on Computer Systems},
  publisher = {ACM},
  author = {Lozi,  Jean-Pierre and Lepers,  Baptiste and Funston,  Justin and Gaud,  Fabien and Quéma,  Vivien and Fedorova,  Alexandra},
  year = {2016},
  month = apr,
  pages = {1–16},
  collection = {EuroSys ’16}
}

@inproceedings{kivity2007kvm,
  author    = {Avi Kivity and Yaniv Kamay and Dor Laor and Uri Lublin and Anthony Liguori},
  title     = {{kvm: The Linux Virtual Machine Monitor}},
  booktitle = {Proceedings of the 2007 Linux Symposium},
  year      = {2007},
  pages     = {225--230},
  url       = {https://www.kernel.org/doc/mirror/ols2007v1.pdf#page=225}
}

@inproceedings{bienia2008parsec,
  series = {PACT ’08},
  title = {The PARSEC benchmark suite: characterization and architectural implications},
  url = {http://dx.doi.org/10.1145/1454115.1454128},
  DOI = {10.1145/1454115.1454128},
  booktitle = {Proceedings of the 17th international conference on Parallel architectures and compilation techniques},
  publisher = {ACM},
  author = {Bienia,  Christian and Kumar,  Sanjeev and Singh,  Jaswinder Pal and Li,  Kai},
  year = {2008},
  month = oct,
  pages = {72–81},
  collection = {PACT ’08}
}

@inproceedings{kasture2016tailbench,
  title = {Tailbench: a benchmark suite and evaluation methodology for latency-critical applications},
  url = {http://dx.doi.org/10.1109/IISWC.2016.7581261},
  DOI = {10.1109/iiswc.2016.7581261},
  booktitle = {2016 IEEE International Symposium on Workload Characterization (IISWC)},
  publisher = {IEEE},
  author = {Kasture,  Harshad and Sanchez,  Daniel},
  year = {2016},
  month = sep 
}

@inproceedings{fasterChandramouli2018,
  series = {SIGMOD/PODS ’18},
  title = {FASTER: A Concurrent Key-Value Store with In-Place Updates},
  url = {http://dx.doi.org/10.1145/3183713.3196898},
  DOI = {10.1145/3183713.3196898},
  booktitle = {Proceedings of the 2018 International Conference on Management of Data},
  publisher = {ACM},
  author = {Chandramouli,  Badrish and Prasaad,  Guna and Kossmann,  Donald and Levandoski,  Justin and Hunter,  James and Barnett,  Mike},
  year = {2018},
  month = may,
  pages = {275–290},
  collection = {SIGMOD/PODS ’18}
}

@MISC{specjbb,
    author = {{Standard Performance Evaluation Corporation}},
    title = {{SPECjbb 2015 benchmark}},
    url  = "https://www.spec.org/jbb2015/",
    date = {2025-08-01},
}

@inproceedings{shi2011limiting,
  title = {Limiting Cache-based Side-channel in Multi-tenant Cloud Using Dynamic Page Coloring},
  url = {http://dx.doi.org/10.1109/DSNW.2011.5958812},
  DOI = {10.1109/dsnw.2011.5958812},
  booktitle = {2011 IEEE/IFIP 41st International Conference on Dependable Systems and Networks Workshops (DSN-W)},
  publisher = {IEEE},
  author = {Shi,  Jicheng and Song,  Xiang and Chen,  Haibo and Zang,  Binyu},
  year = {2011},
  month = jun,
  pages = {194–199}
}

@inproceedings{volos2024principled,
  series = {CCS ’24},
  title = {Principled Microarchitectural Isolation on Cloud CPUs},
  url = {http://dx.doi.org/10.1145/3658644.3690183},
  DOI = {10.1145/3658644.3690183},
  booktitle = {Proceedings of the 2024 on ACM SIGSAC Conference on Computer and Communications Security},
  publisher = {ACM},
  author = {Volos,  Stavros and Fournet,  Cédric and Hofmann,  Jana and K\"{o}pf,  Boris and Oleksenko,  Oleksii},
  year = {2024},
  month = dec,
  pages = {183–197},
  collection = {CCS ’24}
}

@inproceedings{yan2019attack,
  title = {Attack Directories,  Not Caches: Side Channel Attacks in a Non-Inclusive World},
  url = {http://dx.doi.org/10.1109/SP.2019.00004},
  DOI = {10.1109/sp.2019.00004},
  booktitle = {2019 IEEE Symposium on Security and Privacy (SP)},
  publisher = {IEEE},
  author = {Yan,  Mengjia and Sprabery,  Read and Gopireddy,  Bhargava and Fletcher,  Christopher and Campbell,  Roy and Torrellas,  Josep},
  year = {2019},
  month = may,
  pages = {888–904}
}

@inproceedings{hofmann2024gaussian,
  series = {CCS ’24},
  title = {Gaussian Elimination of Side-Channels: Linear Algebra for Memory Coloring},
  url = {http://dx.doi.org/10.1145/3658644.3690263},
  DOI = {10.1145/3658644.3690263},
  booktitle = {Proceedings of the 2024 on ACM SIGSAC Conference on Computer and Communications Security},
  publisher = {ACM},
  author = {Hofmann,  Jana and Fournet,  Cédric and K\"{o}pf,  Boris and Volos,  Stavros},
  year = {2024},
  month = dec,
  pages = {2799–2813},
  collection = {CCS ’24}
}

@inproceedings{kessous2024prune+,
  title = {Prune+PlumTree - Finding Eviction Sets at Scale},
  url = {http://dx.doi.org/10.1109/SP54263.2024.00173},
  DOI = {10.1109/sp54263.2024.00173},
  booktitle = {2024 IEEE Symposium on Security and Privacy (SP)},
  publisher = {IEEE},
  author = {Kessous,  Tom and Gilboa,  Niv},
  year = {2024},
  month = may,
  pages = {3754–3772}
}

@article{kocher2020spectre,
  title = {Spectre Attacks: Exploiting Speculative Execution},
  volume = {63},
  ISSN = {1557-7317},
  url = {http://dx.doi.org/10.1145/3399742},
  DOI = {10.1145/3399742},
  number = {7},
  journal = {Communications of the ACM},
  publisher = {Association for Computing Machinery (ACM)},
  author = {Kocher,  Paul and Horn,  Jann and Fogh,  Anders and Genkin,  Daniel and Gruss,  Daniel and Haas,  Werner and Hamburg,  Mike and Lipp,  Moritz and Mangard,  Stefan and Prescher,  Thomas and Schwarz,  Michael and Yarom,  Yuval},
  year = {2020},
  month = jun,
  pages = {93–101}
}

@inproceedings{oren2015spy,
  series = {CCS’15},
  title = {The Spy in the Sandbox: Practical Cache Attacks in JavaScript and their Implications},
  url = {http://dx.doi.org/10.1145/2810103.2813708},
  DOI = {10.1145/2810103.2813708},
  booktitle = {Proceedings of the 22nd ACM SIGSAC Conference on Computer and Communications Security},
  publisher = {ACM},
  author = {Oren,  Yossef and Kemerlis,  Vasileios P. and Sethumadhavan,  Simha and Keromytis,  Angelos D.},
  year = {2015},
  month = oct,
  pages = {1406–1418},
  collection = {CCS’15}
}

@inproceedings {saileshwar2021mirage,
author = {Gururaj Saileshwar and Moinuddin Qureshi},
title = {{MIRAGE}: Mitigating {Conflict-Based} Cache Attacks with a Practical {Fully-Associative} Design},
booktitle = {30th USENIX Security Symposium (USENIX Security 21)},
year = {2021},
isbn = {978-1-939133-24-3},
pages = {1379--1396},
url = {https://www.usenix.org/conference/usenixsecurity21/presentation/saileshwar},
publisher = {USENIX Association},
month = aug
}

@inproceedings{shahrad2021provisioning,
  series = {SoCC ’21},
  title = {Provisioning Differentiated Last-Level Cache Allocations to VMs in Public Clouds},
  url = {http://dx.doi.org/10.1145/3472883.3487006},
  DOI = {10.1145/3472883.3487006},
  booktitle = {Proceedings of the ACM Symposium on Cloud Computing},
  publisher = {ACM},
  author = {Shahrad,  Mohammad and Elnikety,  Sameh and Bianchini,  Ricardo},
  year = {2021},
  month = nov,
  pages = {319–334},
  collection = {SoCC ’21}
}

@inproceedings{xiang2018dcaps,
  series = {EuroSys ’18},
  title = {DCAPS: Dynamic Cache Allocation with Partial Sharing},
  url = {http://dx.doi.org/10.1145/3190508.3190511},
  DOI = {10.1145/3190508.3190511},
  booktitle = {Proceedings of the Thirteenth EuroSys Conference},
  publisher = {ACM},
  author = {Xiang,  Yaocheng and Wang,  Xiaolin and Huang,  Zihui and Wang,  Zeyu and Luo,  Yingwei and Wang,  Zhenlin},
  year = {2018},
  month = apr,
  pages = {1–15},
  collection = {EuroSys ’18}
}

@inproceedings{xu2018dcat,
  series = {EuroSys ’18},
  title = {dCat: Dynamic Cache Management for Efficient,  Performance-sensitive Infrastructure-as-a-Service},
  url = {http://dx.doi.org/10.1145/3190508.3190555},
  DOI = {10.1145/3190508.3190555},
  booktitle = {Proceedings of the Thirteenth EuroSys Conference},
  publisher = {ACM},
  author = {Xu,  Cong and Rajamani,  Karthick and Ferreira,  Alexandre and Felter,  Wesley and Rubio,  Juan and Li,  Yang},
  year = {2018},
  month = apr,
  pages = {1–13},
  collection = {EuroSys ’18}
}

@inproceedings{vila2019theory,
  title = {Theory and Practice of Finding Eviction Sets},
  url = {http://dx.doi.org/10.1109/SP.2019.00042},
  DOI = {10.1109/sp.2019.00042},
  booktitle = {2019 IEEE Symposium on Security and Privacy (SP)},
  publisher = {IEEE},
  author = {Vila,  Pepe and Kopf,  Boris and Morales,  Jose F.},
  year = {2019},
  month = may,
  pages = {39–54}
}

@article{park2020page,
  title = {Page Reusability-Based Cache Partitioning for Multi-Core Systems},
  volume = {69},
  ISSN = {2326-3814},
  url = {http://dx.doi.org/10.1109/TC.2020.2968066},
  DOI = {10.1109/tc.2020.2968066},
  number = {6},
  journal = {IEEE Transactions on Computers},
  publisher = {Institute of Electrical and Electronics Engineers (IEEE)},
  author = {Park,  Jiwoong and Yeom,  Heonyoung and Son,  Yongseok},
  year = {2020},
  month = jun,
  pages = {812–818}
}

@inproceedings{zhao2024last,
  series = {ASPLOS ’24},
  title = {Last-Level Cache Side-Channel Attacks Are Feasible in the Modern Public Cloud},
  url = {http://dx.doi.org/10.1145/3620665.3640403},
  DOI = {10.1145/3620665.3640403},
  booktitle = {Proceedings of the 29th ACM International Conference on Architectural Support for Programming Languages and Operating Systems,  Volume 2},
  publisher = {ACM},
  author = {Zhao,  Zirui Neil and Morrison,  Adam and Fletcher,  Christopher W. and Torrellas,  Josep},
  year = {2024},
  month = apr,
  pages = {582–600},
  collection = {ASPLOS ’24}
}

@inproceedings {funaro2016ginseng,
author = {Liran Funaro and Orna Agmon Ben-Yehuda and Assaf Schuster},
title = {Ginseng: {Market-Driven} {LLC} Allocation},
booktitle = {2016 USENIX Annual Technical Conference (USENIX ATC 16)},
year = {2016},
isbn = {978-1-931971-30-0},
address = {Denver, CO},
pages = {295--308},
url = {https://www.usenix.org/conference/atc16/technical-sessions/presentation/funaro},
publisher = {USENIX Association},
month = jun
}

@MISC{corescheduling,
    author = {{Linux Kernel Documentation}},
    title = {{Core Scheduling}},
    url  = "https://docs.kernel.org/admin-guide/hw-vuln/core-scheduling.html",
    date = {2025-08-01},
}

@MISC{eevdf,
    author = {{Linux Kernel Documentation}},
    title = {{EEVDF Scheduler}},
    url  = "https://docs.kernel.org/scheduler/sched-eevdf.html",
    date = {2025-08-01},
}

@inproceedings {yang2023skypilot,
author = {Zongheng Yang and Zhanghao Wu and Michael Luo and Wei-Lin Chiang and Romil Bhardwaj and Woosuk Kwon and Siyuan Zhuang and Frank Sifei Luan and Gautam Mittal and Scott Shenker and Ion Stoica},
title = {{SkyPilot}: An Intercloud Broker for Sky Computing},
booktitle = {20th USENIX Symposium on Networked Systems Design and Implementation (NSDI 23)},
year = {2023},
isbn = {978-1-939133-33-5},
address = {Boston, MA},
pages = {437--455},
url = {https://www.usenix.org/conference/nsdi23/presentation/yang-zongheng},
publisher = {USENIX Association},
month = apr
}

@inproceedings{abel2020nanobench,
  title = {nanoBench: A Low-Overhead Tool for Running Microbenchmarks on x86 Systems},
  url = {http://dx.doi.org/10.1109/ISPASS48437.2020.00014},
  DOI = {10.1109/ispass48437.2020.00014},
  booktitle = {2020 IEEE International Symposium on Performance Analysis of Systems and Software (ISPASS)},
  publisher = {IEEE},
  author = {Abel,  Andreas and Reineke,  Jan},
  year = {2020},
  month = aug,
  pages = {34–46}
}

@inproceedings{liu2023cps,
  series = {ASPLOS ’23},
  title = {CPS: A Cooperative Para-virtualized Scheduling Framework for Manycore Machines},
  url = {http://dx.doi.org/10.1145/3623278.3624762},
  DOI = {10.1145/3623278.3624762},
  booktitle = {Proceedings of the 28th ACM International Conference on Architectural Support for Programming Languages and Operating Systems,  Volume 4},
  publisher = {ACM},
  author = {Liu,  Yuxuan and Xu,  Tianqiang and Mi,  Zeyu and Hua,  Zhichao and Zang,  Binyu and Chen,  Haibo},
  year = {2023},
  month = mar,
  pages = {43–56},
  collection = {ASPLOS ’23}
}

@MISC{sysbench,
    author = {{Alexey Kopytov}},
    title = {{Sysbench: scriptable database and system performance benchmark}},
    howpublished = "\url{https://github.com/akopytov/sysbench}",
    date = {2025-08-01},
}

@article{reese2008nginx,
  author       = {Will Reese},
  title        = {Nginx: the High‑Performance Web Server and Reverse Proxy},
  journal      = {Linux Journal},
  year         = {2008},
  month        = sep,
  url          = {https://linuxjournal.com/article/10108}
}

@MISC{kernbench,
    author = {{Linux Test Project}},
    title = {{Kernbench v0.42}},
    url = "https://github.com/linux-test-project/ltp/tree/master/utils/benchmark/kernbench-0.42",
    date = {2025-08-01}
}

@inproceedings{morgan2025slice+,
  title = {Slice+Slice Baby: Generating Last-Level Cache Eviction Sets in the Blink of an Eye},
  url = {http://dx.doi.org/10.1109/SP61157.2025.00264},
  DOI = {10.1109/sp61157.2025.00264},
  booktitle = {2025 IEEE Symposium on Security and Privacy (SP)},
  publisher = {IEEE},
  author = {Morgan,  Bradley and Horowitz,  Gal and O’Connell,  Sioli and van Schaik,  Stephan and Chuengsatiansup,  Chitchanok and Genkin,  Daniel and Maennel,  Olaf and Montague,  Paul and Ronen,  Eyal and Yarom,  Yuval},
  year = {2025},
  month = may,
  pages = {3479–3496}
}

@MISC{schedext,
    author = {{Jonathan Corbet}},
    title = {{The extensible scheduler class}},
    url  = "https://lwn.net/Articles/922405/",
    date = {2025-08-01},
}

@MISC{pbzip2,
    author = {{Jeff Gilchrist}},
    title = {{Parallel BZIP2 (PBZIP2)}},
    url  = "https://compression.great-site.net/pbzip2/",
    date = {2025-08-01},
}

\end{document}